\documentstyle[emulateapj,apjfonts]{article}

\begin{document}   
\def\aph{\alpha_{\rm ph}}   
\def\tobs{t_{\rm obs}}   
\def\Ga{\Gamma}   
\def\be{\begin{equation}}   
\def\cb{\cal B}   
\def\cd{{\cal D}}   
\def\cf{\cal F}   
\def\ee{\end{equation}}   
\def\esyn{\epsilon_{\nu}^{\rm syn}}   
\def\muobs{\mu_{\rm obs}}   
\def\ngt{N(\gamma,t)}   
\def\rp{r^{\prime}}   
\def\rsh{r_{\rm sh}}   
\def\tltr{\tau_{\rm ltr}}   
\def\tesc{\tau_{\rm esc}}   
\def\texp{\tau_{\rm exp}}   
\def\tlag{\tau_{\rm lag}}   
\def\tp{t^{\prime}}   
\def\tsyn{\tau_{\rm syn}}   
\def\tesc{\tau_{\rm esc}}   
   
\title{BLOB EJECTION FROM ADVECTION-DOMINATED ACCRETION FLOW II:   
THE MULTIWAVELENGTH PROPERTIES OF LIGHT CURVES}   
   
\author{Jian-Min Wang\altaffilmark{1,2},
        Masaaki Kusunose\altaffilmark{3}}
	    
\altaffiltext{1}{Laboratory for High Energy Astrophysics, Institute of High Energy    
Physics, The Chinese Academy of Sciences, Beijing 10039, P.R.  China;
wangjm@astrosv1.ihep.ac.cn}
\altaffiltext{2}{School of Physics and Astronomy,   
Tel-Aviv University, Tel-Aviv 69978, Israel}   
\altaffiltext{3}{Department of Physics, School of Science, Kwansei Gakuin University,   
Nishinomiya 662-8501, Japan; kusunose@kwansei.ac.jp}   
   
\slugcomment{submitted to Astrophysical Journal on 27 July/Accepted on 2 Sptember, 2001}

\begin{abstract}   
   
It has been argued that blobs ejected from advection-dominated accretion    
flow through the accretion-ejection instability undergo expansion due to  
their high internal energy density. The expanding blobs interact with  
their surroundings and form strong shock, which accelerates a group of  
electrons to be relativistic. Then flares are formed.  This model has  
advances in two aspects: shock acceleration and self-consistent injection.   
We derive an analytical formula of the injection function of relativistic    
electrons based on the Sedov's law. We calculate the time-dependent spectrum  
of relativistic electrons in such an expanding blob. The light-travel effect,  
the evolution of the electron spectrum due to energy loss, and the escape  
of relativistic electrons from the radiating region are considered, as well  
as the expansion (at sub-relativistic speed) of the coasting blob.    
A large number of light curves spanning wide spaces of parameters   
have been given in this paper.    
Regarding the symmetry, relative amplitude, duration of a flare, and the   
time lag between peak fluxes, we find four basic kinds of light curves for    
the non-expanding blob, and seven basic    
kinds of light curves for the expanding blob.  The expansion weakens   
the magnetic field as well as enlarges the size of the blob.   
The predicted light curves thus show very complicated properties   
which are composed of the basic light curves, because the physical conditions   
are changing in the expanding blob. For the rapid decays of magnetic  
filed (for example, $B \propto r^{-n}$), we find the falling profile is a  
power law of time as $\nu F_{\nu}\propto t^{(1-\alpha)n/5}$,  
where $\alpha$ is the power-law index of injected electrons.  
This decline is controlled by the decay of magnetic field rather than the  
energy losses of relativistic electrons.  We also calculate the evolution   
of the photon spectrum from both non-expanding and expanding blobs.    
Different shapes in the phase of decreasing luminosity are then obtained for   
different parameter values.   
The photon index, $\alpha_{\rm ph}$, keeps constant for non-expanding   
blobs when luminosity decreases, whereas $\alpha_{\rm ph}$ continues to  
decrease after the luminosity reaches its maximum for expanding blobs.   
It is expected that we can extract the information of ejected blobs from    
the observed light curves based on the present model.   
   
\keywords{radiation mechanism-synchrotron-quasar-blazar}   
\end{abstract}   
   
\section{Introduction}   
   
The time variations of received fluxes and emission spectra of    
blazar-type active galactic nuclei (AGNs) have received great attention   
from both theory and observation during the last 30 years,   
because the multiwavelength observations of flares from blazars provide   
invaluable information how the central engines work (Bregman 1990;   
Urry \& Padovani 1995; Ulrich et al. 1997).  The flare variabilities from    
blazars are very common.  However, the properties of these flares are diverse,   
and the outbursts of the flares are over multiwavelength.   
Although we have a plenty of observational data of these phenomena,   
our understanding of the mechanisms of the flares are still insufficient.   
   
Many attempts have been made to fit the observed continuum spectra by 
varieties of theoretical models. These calculations can be mainly divided 
into two kinds: homogeneous models (blob models) and inhomogeneous models 
(elongated jet models).  At early stage the homogeneous models were   
frequently invoked owing to their concise physical scenario. Rees (1966, 
1967) and Ozernoy \& Sazonov (1968) originally studied the emergent 
spectrum from an expanding blob disregarding the evolution of the electron 
spectrum and blob dynamics. Subsequently, the expanding process was studied 
(Mathews 1971; Vitello \& Salvati 1976; Canuto \& Tsiang 1977; Christiansen, 
Scott, \& Vestrand 1978). The free expanding case was considered under the 
assumptions that the electron  spectrum is steady without time evolution   
(Jones \& Tobin 1977; Vitello \& Pacini 1977 and 1978;   
K\"onigl 1978; Marscher 1980; Band \& Grindlay 1986).   
Then the effect of electron spectrum evolution with presumed injection   
function was generally considered (Celotti, Maraschi, \& Treves 1991;   
Wang \& Zhou 1995; Mastichiadis \& Kirk 1997; Kirk, Rieger, \& Mastichiadis 
1998; Dermer, Sturner, \& Schlickeiser 1997; Dermer 1998; Georganopoulos \& 
Marscher 1998; Chiaberge \& Ghisellini 1999; Li \& Kusunose 2000;    
Kusunose, Takahara, \& Li 2000). The random fluctuations of the continuum 
from blazars are generally attributed to the propagation of relativistic shock    
(Marscher \& Gear 1985).  Large flares most probably result from a blob   
(Dermer, Sturner, \& Schlickeiser 1997; Chiaberge \& Ghisellini 1999)    
as considered by the authors listed above. It should be cautioned that all 
the calculations avoid three important processes: 1) how to accelerate the 
electrons in the blob, 2) how to inject the accelerated electrons into the 
blob if radiation regions are separated from acceleration regions, and 3) 
why and how to eject a blob. It is generally believed that these problems are 
related with accretion disks around supper massive black holes as a central 
engine. The status of accretion disks may determine the properties of the 
ejected blobs.   
   
The properties of the iron K$\alpha$ line emission in radio-loud quasars   
may provide a new clue to understand the flare process, because it is a 
powerful probe to test the ionization state of matter close to black holes. 
The available data show that the line profiles in radio-loud quasars are 
very narrow (Sambruna et al. 1999; Reeves \& Turner 2000), which are quite 
different from radio-quiet quasars and Seyfert galaxies. Thus an optically 
thin advection-dominated accretion flow (ADAF, Narayan \& Yi 1994) or 
ion-pressure supported tori (Rees et al. 1982) may power the central engine 
in radio-loud quasars, which may be explained by the properties of iron 
K$\alpha$ line (Sambruna et al. 1999). If the accretion powers the central 
engine, then the accretion rate is roughly of order   
\begin{equation}   
\dot{m} \sim 10^{-3}\cd_{10}^{-4}L_{47}M_8 \, ,   
\end{equation}   
where $\dot{m}$ is the dimensionless accretion rate normalized by the Eddington   
rate, $\cd_{10}$ is the Doppler factor normalized by 10, i.e.,   
$\cd_{10} = \cd/10$, $L_{47}$ is the bolometric luminosity    
normalized by $10^{47}$ erg/s, i.e.,   
$L_{47} = L_{\rm bol}/10^{47}$, and $M_{8}$ is the black hole mass    
normalized by $10^8 M_{\sun}$, i.e.,    
$M_8 = M_{\rm BH}/(10^8 M_{\sun})$. Such a crude estimation is in  
agreement with that $\dot{m}\sim 10^{-2\sim -3}$ in FRII galaxies evaluated by   
Ghisellini \& Celotti (2001).  
Indeed, such a low accretion rate implies that optically thin ADAF or    
ion-pressure supported tori may power the central engine.    
The observed bolometric luminosity suggests that the emission regions are   
highly relativistically boosted, which dominates   
over the emission from accretion disks (or ion-pressure supported tori).   
    
Tagger \& Pellat (1999) showed there is a strong accretion-ejection    
instability, which increases with the height of an accretion disk.    
Based on this instability, Wang et al. (2000) suggested that    
there is a large flare if a blob ejection takes place in optically thin    
ADAF of blazars.  Comparing with the blob's surroundings,    
Wang et al. (2000) also suggested the observational consequences of    
an ejected blob: nonthermal flares and recombination line emission,    
which are caused by the expansion of a blob owing to its high internal    
energy density. In fact there is an ample of evidence for the blob ejection 
from the nucleus. Especially, recent observations of 3C 273 by {\it Chandra}   
discovered the blob ejection from the central region (Marshall et al. 2001).   
If the flare arises from blob ejection, according to the model of Wang et al. 
(2000), what are the properties of the ejected blob in multiwavelength continuum?   
It is the goal of our paper to calculate the multiwavelength light curves    
of the emission from the ejected blob.    
   
This paper is arranged as follows.  Section 2 describes   
our model and the formulation, and an analytical injection function is    
derived from Sedov's self-similar expansion.    
Extensive qualitative analyses of light curves are given in \S 3.    
To compare with observations we provide a large number   
of light curves and confirm the qualitative analysis in \S 4.   
Discussions and conclusions are given in the last section.   
     
\section{The descriptions of models}   
   
Supposing that the inter-cloud medium is homogeneous, the expansion    
of a blob is quite similar to the $\gamma$-ray burst fireball.   
As shown by Wang et al. (2000) the expansion is sub-relativistic.   
The expansion leads to the formation of strong shock.   
Thus such an expansion can be described by the self-similar solution after   
the timescale $t_1$ (Sedov 1969), which is rather shorter than that   
of the expansion phase.  Sedov (1969) gives the self-similar solution:   
the shock speed $v_{\rm sh}\propto t^{-3/5}$ and shell radius   
$r_{\rm sh}\propto t^{2/5}$. Some electrons in the swept medium by   
blob expansion will be accelerated to be relativistic.  Therefore, the   
expansion acts as the injection of relativistic electrons.   
We follow the usual assumption that a power-law distribution 
($\gamma^{-\alpha}$) of electrons is formed via shock acceleration, 
where $\gamma$ is the Lorentz factor of an electron.   
The acceleration timescale is assumed to be much shorter than the   
expansion and synchrotron timescales (for the acceleration timescale, e.g.,  
Kirk et al. 1994 and references therein).  
Then the injection function $Q(\gamma, t)$ can be written as    
$Q(\gamma,t) \propto r_{\rm sh}^2 v_{\rm sh}\propto t^{1/5}$.   
Because the expansion of the blob will be drastically decelerated   
after long enough time, we introduce a cutoff factor $e^{-t/t_2}$,   
\begin{equation}   
\label{eq:qinj}   
Q(\gamma,t) = Q_0\gamma^{-\alpha}t^{1/5}e^{-t/t_2} \, ,   
\end{equation}   
where $Q_0$ can be obtained by the total injected energy $E_0$ after   
integrating the time and electron energy,  
\begin{equation}   
\label{eq:q0}   
Q_0\approx \frac{6(2-\alpha)E_0}{5m_ec^2   
            \left[(\gamma^0_{\rm max})^{2-\alpha}-   
          (\gamma^0_{\rm min})^{2-\alpha}\right]t_2^{6/5}}, 
          ~~~{\rm for}~~\alpha\ne 2\, .   
\end{equation}    
For $\alpha = 2$, we have $Q_0 \approx 6 E_0/[5m_ec^2t_2^{6/5}   
\ln (\gamma_{\rm max}/\gamma_{\rm min})]$. Equation (2) implies that 
the injection is roughly constant within a timescale $t_2$.    
The evolution of electron energy can be determined by   
$\gamma = \gamma_0/(1+a\gamma_0t)$, where $\gamma_0$ is the initial    
energy of electron and $a = c\sigma_{\rm T} B^2/6\pi m_ec^2=1.29\times 10^{-9}B^2$.  
Here $c$ is the light speed, $\sigma_{\rm T}$ is the Thomson cross section,   
$m_e$ is the electron mass, and $B$ is the magnetic field strength in units of Gauss.    
The electron continuity equation in the homogeneous expanding blob reads   
\begin{equation}   
\frac{\partial \ngt}{\partial t}+\frac{\partial}{\partial \gamma}    
                    \left[b(\gamma)\ngt\right]   
                    +\frac{\ngt}{\tau_{\rm esc}} = Q(\gamma,t) \, ,   
\end{equation}   
where $b(\gamma)=-a\gamma^2$    
and $\tau_{\rm esc}$ is the timescale of electron escape from the   
radiation region, which determines the later evolution of   
a flare showing up in lower frequencies.  If re-acceleration of relativistic   
electrons takes place the pile-up distribution will be formed in the shell    
(Schlickeiser 1984), but we neglect the re-acceleration here.    
Therefore the solution of the continuity equation of electrons is given by   
(Ginzburg \& Syrovatskii 1965)   
\begin{equation}   
\label{eq:el-dist}   
N(\gamma,t)=\frac{1}{b(\gamma)} \int_{\gamma}^{\gamma_{\rm max}}   
     \exp\left(-\frac{\tau}{\tau_{\rm esc}}\right)   
     Q(\gamma_0,t-\tau) d\gamma_0 \, ,   
\end{equation}   
where   
$\tau = \int_{\gamma_0}^{\gamma}b^{-1}(\gamma^{\prime})d\gamma^{\prime}   
    = \left(\gamma^{-1}-\gamma_0^{-1}\right)/a$.   
The number density of the electrons reads   
$n(\gamma,t)=N(\gamma,t)/(4\pi r^2_{\rm sh}\Delta r_{\rm sh})$.  These   
equations describe the evolution of the relativistic electrons.   
   
It is known that the observed light curve is ``contaminated''   
by two factors: the size of a blob and the evolution of the relativistic electrons.   
Therefore the information of geometric structure and motion is hidden    
in the light curves as well as the pure evolution of electrons.   
Figure 1 shows the geometry of the expanding blob in its coasting frame.   
Suppose at a given time $t_{\rm blob}$ (in the blob   
frame), the location of its center is $C^{\prime}$,    
and the distance to the observer is $d_{\rm o^{\prime}c^{\prime}}$.   
For the arbitrary point $P(r_{\rm sh}, \theta^{\prime}, \phi^{\prime})$    
on the shell, the distance of $P$ to the observer reads   
\def\dop{d_{_{\small {\rm O^{\prime}P}}}}   
\def\doc{d_{_{\small {\rm O^{\prime}C^{\prime}}}}}   
\begin{equation}   
\dop \approx \doc    
\left[1 -\left(\frac{r_{\rm sh}}{\doc}\right)\cos \psi \right] \, ,  
\end{equation}   
where $\cos \psi=\sin \theta^{\prime} \sin (\phi^{\prime}+\Theta)$,  
$\psi$ is angle PC$^\prime$O$^\prime$, and $\Theta$ is the angle   
between the line of sight and the motion of the blob.  
This approximation is always valid because of  
$d_{\rm o^{\prime}c^{\prime}}\gg r_{\rm sh}$.    
Here we neglect the change of angle $\Theta$ due to the motion of the blob,   
because the distance of the blob is so far away from the observer.   
It is pointed out that we compare the observer time $t_{\rm obs}$   
with the time at the center of the blob $t_b^\prime$, i.e.,  
$t_{\rm obs} = t_b^{\prime} / {\cal D}$, where   
$\cal{D}$ is the Doppler factor given by  
${\cal{D}}=1/\Gamma(1-\beta \cos \Theta)$, with $\Gamma$ being the Lorentz factor of  
the blob motion and $\beta = (1-1/\Gamma^2)^{1/2}$.  
Then we can set up the relation between radiation time $t^{\prime}$ (time at $P$)   
and the blob time at the center:  
\begin{equation}   
\tp = t_{\rm b}^{\prime}+\frac{\doc-\dop}{c}   
  \approx t_{\rm b}^{\prime}+\left(\frac{r_{\rm sh}}{c}\right)\cos \psi \, ,   
\end{equation}   
with sufficient accuracy. The retardation is clear in this equation.   
The shell thickness $\Delta r_{\rm sh}$ increases with the radius of    
the blob, but $\Delta r_{\rm sh}/r_{\rm sh}\ll 1$ holds for the expansion    
phase (Sedov 1969). Thus our treatments are valid.   
   
The synchrotron emissivity of relativistic electrons is denoted   
by $\epsilon_{\nu}$, then the emission from a shell element    
$(\Delta r_{\rm sh},\Delta \theta^{\prime},\Delta \phi^{\prime})$    
should be    
$\Delta L_{\nu}=\epsilon_{\nu}(r^2_{\rm sh} \Delta r_{\rm sh})   
\sin \theta^{\prime} \Delta \theta^{\prime} \Delta \phi^{\prime}$,    
where 
$\epsilon_{\nu}(t)=c_3 B   
\int_{\gamma_{\rm min}}^{\gamma_{\rm max}} n(\gamma,t) F(x) dx$,   
$x = \nu/\nu_0B\gamma^2$,    
$F(x) = x \int_x^{\infty} K_{5/3}(z) dz$, and $K_{5/3}$ is the modified    
Bessel function with order 5/3, $\nu_0=3e/4\pi m_e c = 4.21\times 10^6$,    
and $c_3 =\sqrt{3}e^3/(4 \pi m_e c^2) = 1.87\times 10^{-23}$ (Pacholczyk 1973).    
The blob is optically thin in our interested bands, which are much higher   
than the synchrotron self-absorption frequency.    
The emergent spectrum from the shell of the expanding blob can be written as   
\begin{equation}   
\label{eq:lumi0}  
L(\nu^{\prime},t_{\rm b}^{\prime})   
                   = c_3 B  \int_0^{\pi}\sin \theta^{\prime}    
                   d\theta^{\prime} \int_0^{2\pi}  d\phi^{\prime}   
                      \int_{\gamma_2}^{\gamma_1} H(\zeta)   
                      N(\gamma,\tp) F(x)d\gamma \, ,   
\end{equation}   
where $H(\zeta)$ is the Heaviside function: $H(\zeta)=0$ for $\zeta < 0$,  
whereas $H(\zeta) = 1$ for $\zeta \geq 0$. Here parameter $\zeta$ is given by   
\begin{equation}   
\zeta = t_{\rm b}^{\prime}+\frac{r_{\rm sh}}{c}\cos \psi \ge 0 \, ,   
\end{equation}   
which is incorporated from equations (6) and (7). The introduction   
of the Heaviside function is based on the retardation of light    
propagation.  Because of the infinitesimal thickness of the shell, 
the retardation of photons to the observer is caused only by emission 
from different points ($\theta^\prime$, $\phi^\prime$).   
It is interesting to note that the emission (eq. 8) is independent    
of the specific geometric thickness of the shock.   
This leads to the avoidance of the discussion of the shock thickness.    
   
We assume the magnetic field changes with radius as    
\begin{equation}   
\label{eq:mag-exp}   
B(r)=B_0\left(\frac{r}{\rsh^0}\right)^{-n},   
\end{equation}   
where $\rsh^0$ is the radius of blob at $t=t_1$. The index $n$ remains   
a free parameter in our model. The flux from the whole blob received by 
a distant observer is obtained by Lorentz transformation as   
\begin{equation}   
F({\nu},t_{\rm obs})=\frac{1}{4\pi d_L^2}{\cal D}^3   
                     L\left(\frac{\nu}{\cal{D}},{\cal D} t_{\rm obs} \right)\, ,   
\end{equation}   
where $d_L$ is the luminosity distance from the source to the observer.   
Here we neglect the difference in the directions of emitting locations   
in the blob measured from the line of sight.   
Also for simplicity we neglect the difference of Doppler factor at different   
locations in the blob, because the expansion is sub-relativistic.   
   
As a summary of the parameters employed in the present model,   
all of them are listed in Table 1.  There are 12 parameters describing    
the processes in the blob. Some of them are the initial values of the blob, 
for example, $B_0$, $\rsh^0$, and $\beta_0$. With the help of Sedov 
self-similar solution, these parameters at any time will be known. The 
escape timescale is taken to be a free parameter. The dependence of 
magnetic field on the shell radius owing to expansion is an adjustable 
parameter. The timescale of injection $t_2$ is a new parameter, and $t_1$ 
in fact connects with the initial expansion velocity $\beta_0$ (eq. 23 below).  
For a non-expanding blob there are 11 parameters (1 -- 11).    
The non-expanding blob in fact is homogeneous for a distant observer,    
whereas the expanding blob shows the inhomogeneous properties as we show below.

\section{Qualitative analysis}   
   
Generally speaking there are four aspects to describe the characteristics    
of the light curves at different wavebands: 1) the symmetry of light curves 
about the peak flux, 2) the relative amplitudes of light curves at the peaks, 
3) duration of the bursts, which is defined by the $e$-folding time of 
$\nu F_{\nu}$, 4) the time lag of the light curves, which is defined as   
the time interval between the peak fluxes at different wavebands.    
Sophisticated observations should discover these   
properties by comparing observed spectra with the theoretical models.   
  
From equation (\ref{eq:lumi0}) we know that there are four processes affecting 
the observed light curves: blob's expansion which increases the blob's size, 
energy losses of electrons via synchrotron radiation, photon traveling, and the 
escape of electrons from the blob shell. The four processes are competing each 
other, determining the properties of light curves, and are hidden in the 
observed light curves. We first analyze the roles of the size,  escape, and 
energy losses in the light curves before making detailed numerical calculations.    
   
\def\DE{\Delta E}   
\def\tnu{\tau_{\rm syn}^{\nu_1}}   
\def\tnuu{\tau_{\rm syn}^{\nu_2}}   
   
\subsection{Non-expansion cases}   
   
Considering the complexities of the expanding blobs, we first   
analyze the most simple cases of the non-expansion blobs.    
There are three important timescales, i.e., 1) $\tau_{\rm syn}$: energy loss 
timescale due to synchrotron radiation, which is proportional to $\nu^{-1/2}$, 
2) $\tltr$: light-travel timescale through the blob, 3) $\tesc$: the timescale 
of electron escape from the radiation region. Here we consider two light curves 
of photons with frequencies $\nu_1$ and $\nu_2$ ($\nu_1 > \nu_2$). The lifetimes 
of relativistic electrons which radiate photons at frequencies $\nu_1$ and 
$\nu_2$ are denoted by $\tau_{\rm syn}^{\nu_1}$ and $\tau_{\rm syn}^{\nu_2}$, 
respectively. The total energy  radiated at $\nu$ during lifetime 
$\tau_{\rm syn}^{\nu}\propto \gamma^{-1}$ is given by   
\begin{equation}   
\label{eq:e-loss12}   
\DE \sim \gamma N(\gamma, t)\propto \gamma^{1-\alpha} \, .   
\end{equation}   
For $\nu_1$ and $\nu_2$, their total radiated energies are   
$\DE_1\propto \gamma_1^{1-\alpha}$ and    
$\DE_2\propto \gamma_2^{1-\alpha}$ within $\tnu$ and $\tnuu$, respectively.    
Generally the size of the blob determines the duration of a flare, if $\tltr$ 
is long enough. The escape effect will sharpen the falling profile and the peak 
flux, if $\tesc$ is short enough. The retardation of the peak flux at different 
bands is obviously determined by the above three processes in the blob, i.e., 
synchrotron loss, light travel, and electron escape. The competition among the 
three timescales determines the light curves from the blob.   
According to these timescales, we distinguish the following four kinds   
of light curves for a non-expanding blob.   
   
(NE1) For the case of $\tnu<\tnuu<\tltr\ll \tesc$, energy loss timescales   
are shorter than $\tltr$ and much shorter than $\tesc$.    
This implies that the escape can be neglected.   
Under such a circumstance, only one process, namely, light traveling,   
determines the observed light curves, because the lifetime of electrons    
is shorter than light travel timescale $\tltr$. Then the observed light 
curves are mainly due to the light travel effects. The light curves should 
be symmetric about the peak flux, because the rising time and the falling 
time are roughly equal to $\tltr$. The time lag may roughly be zero.    
The radiation processes of relativistic electrons will be hidden.     
The duration of the flare is roughly equal to light travel timescale $\tltr$, 
too. Now we define LC$_1$ and LC$_2$ as the light curves at frequencies   
$\nu_1$ and $\nu_2$, respectively. The peak fluxes of LC$_1$ and LC$_2$ are 
denoted by $L^{\rm pk}_1$ and $L^{\rm pk}_2$, respectively.   
Thus the ratio of the peak fluxes will be given by   
\begin{equation}   
\label{eq:amp-ne1}   
\frac{L^{\rm pk}_1}{L^{\rm pk}_2} \sim    
\frac{\DE_1/\tltr}{\DE_2/\tltr}=   
         \left(\frac{\gamma_1}{\gamma_2}\right)^{1-\alpha}=   
         \left(\frac{\nu_1}{\nu_2}\right)^{(1-\alpha)/2},   
\end{equation}   
from equation (12).    
   
(NE2) For the case of $\tnu<\tltr<\tnuu\ll \tesc$. The longer timescale   
of escape guarantees the rough conservation of the number of radiating 
electrons in the blob. The light curve at frequency $\nu_1$, LC$_1$,   
is still determined by the blob's geometry, but one at $\nu_2$,    
LC$_2$, will be determined by radiation processes and blob's    
geometry.  The rough symmetry remains in LC$_1$, whereas asymmetry in    
LC$_2$ appears.  This is simply owing to the differences between rising    
timescale (due to light travel and energy losses) and falling timescale    
(due to radiation). The duration time of LC$_1$ will be determined by the    
light travel timescale $\tltr$, and the duration of LC$_2$ will be modulated   
by light travel timescale and radiation timescale $\tltr + \tnuu$.    
The time lag is controlled by synchrotron radiation and light travel.   
The duration of LC$_1$ is still determined by the blob's size,    
but the duration of LC$_2$ will be roughly determined by synchrotron radiation.   
The ratio of the peak flux will be roughly given by   
\begin{equation}   
\label{eq:amp-ne2}   
\frac{L^{\rm pk}_1}{L^{\rm pk}_2} \sim    
\frac{\DE_1/\tltr}{\DE_2/(\tltr+\tnuu)}=   
         \left(1+\frac{\tnuu}{\tltr}\right)   
         \left(\frac{\nu_1}{\nu_2}\right)^{(1-\alpha)/2}\, ,   
\end{equation}   
implying the combined effects of the spectrum, size, and magnetic field.   
   
(NE3) For the case of $\tltr<\tnu<\tnuu\ll \tesc$. The shortest timescale 
is the travel timescale, which means that the blob's geometry is less 
important in the light curves. The time lag is caused by the differences 
of energy-loss timescales at $\nu_1$ and $\nu_2$.  The retardation of the 
flux peak will be of order   
\begin{equation}   
\label{eq:lag-ne3}   
\tlag\sim \tnuu -\tnu \, .   
\end{equation}   
The ratio of the amplitudes is determined by the spectrum    
of relativistic electrons, roughly given by    
\begin{equation}   
\label{eq:amp-ne3}   
\frac{L^{\rm pk}_1}{L^{\rm pk}_2} \sim    
\frac{\DE_1/\tnu}{\DE_2/\tnuu}=   
              \left(\frac{\nu_1}{\nu_2}\right)^{(2-\alpha)/2} \, .   
\end{equation}   
The duration of a flare is mainly determined by the lifetime of    
electrons radiating at this frequency, which can be roughly expressed   
by $\tnu/\tnuu\sim (\nu_2/\nu_1)^{-1/2}$.  The profile of light curves    
may be asymmetric, because the rising timescale is determined by the 
injection, and the falling timescale is determined by energy losses.    
    
(NE4) The case of $\tnu<\tesc<\tnuu<\tltr$. The above three cases   
are presumed that the escape timescale is much longer than any other   
timescales.  This assumption guarantees the conservation of electron number   
during the whole lifetime of relativistic electrons. The escape effects 
result in the decrease of the electron number (see Eq. \ref{eq:el-dist}).     
Equivalently, the total luminosity will be lowered. Now we consider an 
example of electron escape. In such a case the escape and light travel 
effects are important in the light curves. If the escape timescale of 
electrons is shorter than light travel and synchrotron radiation timescales, 
the case will be meaningless in observations, because the light curves are 
mainly determined by the escape process.    
On the other hand, if $\tnu<\tesc<\tnuu<\tltr$ holds,    
the ratio of the amplitudes will be given by   
\begin{equation}   
\label{eq:amp-ne4}   
\frac{L^{\rm pk}_1}{L^{\rm pk}_2} \sim    
\frac{\DE_1/\tnu}{\DE_2/\tesc}=   
         \frac{\tesc}{\tnu}   
         \left(\frac{\nu_1}{\nu_2}\right)^{(1-\alpha)/2}\, ,   
\end{equation}   
which is controlled by the electron spectrum, escape, and magnetic field,   
and this case will be complicated. The rising profile of LC$_2$    
is caused by synchrotron radiation, whereas its falling profile is    
by the escape process. The asymmetry appears in light curves, and the 
relative amplitude will be determined by escape and radiation processes.   
   
The above discussions are summarized in Table 2, which gives the   
properties of light curves according to symmetry, duration, relative   
amplitude, and time lag.   
We rule out the case of $\tesc< \min (\tltr,\tnu,\tnuu)$,   
because electrons will directly escape without emission.   

   
\subsection{Self-similar expansion}   
   
The expansion of a blob causes the light curves more complicated   
because of an additional timescale $\texp$.    
The light curves from an expanding blob will be given by the above    
four characteristic light curves with the modulations by expansion.    
Generally the expansion timescale   
is longer than that of the light travel timescale in the present model,   
because we are considering the non-relativistic expansion.    
Complexities arise from the changes of magnetic field and size of the blob   
due to expansion.  Based on the four kinds of light curves described above,    
it is easy to distinguish the new characteristics of    
light curves from an expanding blob.  We can compare    
the expansion timescale with the other three timescales;    
Table 2 is the summary of the meaningful light curves according to    
the four aspects of symmetry, duration, relative amplitude, and time lag.   
  
The expansion not only enlarges the size of the blob,    
but also weakens the magnetic field.  The enlarging size makes    
the light travel timescale longer;   
the weakening magnetic field dilates timescale of    
synchrotron emission but lowers the output emission power.  From    
equation (\ref{eq:mag-exp}),   
we obtain the synchrotron timescale   
\begin{equation}   
\label{eq:syn-time}   
\tau_{\rm syn}(\gamma)=\tau_{\rm syn}^0(\gamma)   
                           \left(\frac{r}{\rsh^0}\right)^{2n} \, ,   
\end{equation}   
where $\tau_{\rm syn}^0 = 1.0 \times 10^{9}/(\gamma B_0^2)$ s,   
with $\rsh^0$ and $B_0$ being the initial radius and magnetic field of    
the blob, respectively.   
For a mono-energetic electron with energy $\gamma$,    
the energy loss rate is given by   
\begin{equation}   
\label{eq:dotgamma}   
\dot{\gamma}(\gamma)=\dot{\gamma}_0(\gamma)   
                     \left(\frac{r}{\rsh^0}\right)^{-2n},   
\end{equation}   
and the peak frequency of its emission is   
\begin{equation}   
\label{eq:nuc}   
\nu_{\rm c}(\gamma)=\nu_{\rm c}^0(\gamma)\left(\frac{r}{\rsh^0}\right)^{-n},   
\end{equation}   
where $\dot{\gamma}_0=1.29\times 10^{-9}B_0^2\gamma^2$   
and $\nu_{\rm c}(\gamma)=4.21 \times 10^{6}\gamma^2B_0$ Hz.     
The index $n$ is between 1 and 2 for different models.  From this equation    
we see that the synchrotron timescale is very sensitive to    
the expansion law of a blob. This makes the light curves very complicated, 
showing more than one kind  of light curves in one flare.     
   
Although the light curves behave similarly to those of the non-expanding blob, 
four timescales such as $\tsyn$, $\tltr$, $\texp$, and $\tesc$ jointly control  
the light curves. Comparing them for the expanding blob, we obtain seven 
kinds of light curves and list them in Table 2 according to their basic  
properties such as symmetry, duration, relative amplitude and time lag.  
   
\section{QUANTITATIVE ANALYSIS}   
   
In this section we show our numerical results.     
As we already stressed, the present model includes the expansion effects through    
two processes of enlarging size and weakening magnetic field. We first show the 
results of a non-expanding blob, and then show the results of several cases for 
different parameters of an expanding blob, and last we compare the results of 
non-expanding and expanding blobs. We list all the cases for non-expansion and 
expansion in Tables 3 and 4, respectively. The photon energies are taken to be 
0.1, 1, 10, 100 eV and 1, 2, 3, 4 keV, which span almost 5 orders of magnitude 
from near infrared, optical, UV, EUV, soft X-ray, to the middle range of X-ray 
bands. The Doppler factor is taken to be ${\cd} = 10$,    
and the total injection energy $E_0 = 10^{45}$ erg within time interval $t_2$.   
Also $\gamma_{\rm min} = 10^2$ and $\gamma_{\rm max} = 5 \times 10^5$ are 
assumed in all models. Finally, $\Gamma = 1/ \Theta$ is assumed.  
   
\subsection{Electron energy spectra}   
   
Taking the form of equation (\ref{eq:qinj}) as the injection function, the 
electron energy spectrum can be obtained from equation (\ref{eq:el-dist}).   
Here we should note that the injection function is based on Sedov's self-similar 
solution and the evolution of the electron spectrum is self-consistent.    
Figures 2a and 2b show the electron spectra for two cases.    
For the non-expanding case, the number density increases with time for $t < t_2$.  
The increase is linear, because the injection rate is almost constant before time 
$t_2$.  Another prominent characteristic is the break at high energy.  This is 
due to cooling of synchrotron radiation. The difference in the spectral index 
between the low and high energy is unity. This will be reflected in the photon 
spectral energy spectrum as we show in the following figures of photon index 
evolution.  
     
There are two possible cases for expansion cases.   
1) The injection is faster than the energy losses.  Since the synchrotron  
timescale is dependent on the electron energy, this condition will be satisfied 
for electrons with high energy in the late phase of the expansion by considering  
the magnetic field decays with time as $t^{2n/5}$.  In such a case the number  
density will almost linearly increase with time. 2) Synchrotron loss is more 
rapid than expansion ($\tsyn < \texp$). Then the number density of high energy 
electrons will decrease, because the injection is too slow. Such a case may 
appear in the early phase of expansion, because the condition ($\tsyn < \texp$) 
will be satisfied.  Then the spectral break will appear.  
In Figure 2b, the above two characteristics are evident:   
in the early phase there is a break due to energy losses,  
whereas in late phase the number density increases linearly  
and the break becomes weak. { This will result in different behaviors 
in the plot of spectrum index versus luminosity in the later phase.}  
      
\subsection{Non-expanding Blob}   
   
We now consider a non-expanding shocked shell of the blob in order to   
facilitate to compare with the expanding situations, although the non-expanding 
model may not operate in practice. Here we assume that the injection still 
takes the form of equation (\ref{eq:qinj}). The size and magnetic field hold 
constant. Therefore nine parameters are in the non-expansion model.  
Except $t_2$, other eight parameters are often used  
in homogeneous model, namely, the magnetic field   
$B$, Doppler factor ${\cal D}$, the injected total energy $E_0$, the   
index of the injected electron spectrum, the maximum and minimum Lorentz   
factors of the electron population, the size of the blob $\rsh^0$, and   
the last is the timescale of electron escape from the radiation region.   
   
In this model the most important timescale is the lifetime of electrons   
radiating synchrotron photons with $\nu$.   
The simple estimation of the lifetime is given by   
\begin{equation}   
\label{eq:tausyn}   
\tau_{\rm syn}=2.0\times 10^4 B^{-3/2}{\cal D}^{-1/2}\nu_{15}^{-1/2}~ {\rm s},   
\end{equation}   
where all the parameters are in observer's frame, and $\nu_{15}=\nu/10^{15}$.   
The lifetime is insensitive to Doppler factor. We always keep the Doppler factor 
constant in various models. In this case the duration time, time lag, and 
relative amplitude of a flare in different bands will be independent of the 
total injected energy. Although there are other seven parameters in the simple 
homogeneous model, the key parameters are $\rsh^0$, $B$, and $\tau_{\rm esc}$ 
which determine the duration time, the time lag, and the relative amplitudes 
in different bands. We show the details in Figs 3 -- 7.   
      
\subsubsection{Size effects}   
   
Size effects are shown in Figure 3.  We change the size as a free parameter and 
fix all other parameters. We take the size of the blob, $\rsh$, to be $10^{15}$,   
$10^{16}, 5 \times 10^{16}$, and $10^{17}$ cm to find the effects of    
the size on light curves.     
   
{\it Symmetry:} The lowest panel in Figure 3 shows the light curves    
for $\rsh=10^{17}$cm.  It is interesting to find that the light    
curves in high energy bands (right panel) are symmetric about the peak,    
whereas the rising timescale is much shorter than the falling    
timescale in the low energy band (left panel).  This quantitatively    
confirms the qualitative analysis in \S 3.  The rising and falling    
timescales are determined by the light travel timescale $\tltr$ as we    
listed for high energy light curves in Table 2.  This characteristic    
can be justified by comparing $\tltr$ and $\tsyn$.  From the left panel,   
we see that the symmetry of light curves is gradually broken with the decrease 
of the size of the blob.  We also see that the rising timescale decreases    
rapidly with $\rsh$, because it is mainly determined by $\max(\tltr,\tsyn)$.  
Usually the rising time is shorter than the falling timescale for the asymmetric 
light curves. Other asymmetric light curves can be found   
in Figure 3 according to the condition listed in Table 2.   
   
\def\tD{\tau_{\rm d}}   
{\it Duration} ($\tD$): We find the duration time is very sensitive to   
the size of a blob, because we defined the duration time as the $e$-folding  
time of $\nu F_\nu$.  
Generally $\tD$ will be prolonged by the size.  The duration   
time can be roughly given by $\tD\sim \max (\tsyn,\tltr)$.   
The keV photons are radiated from electrons with high energy,    
and it is expected that $\tsyn < \tltr$.  Thus $\tD$ will be given    
by $\tltr$.  From the figures in the left panel we find that $\tD$ of    
lower energy photons is much longer than those in the right panel.    
The lower energy photons produced by electrons with lower energy,    
and it is expected that $\tsyn >\tltr$.  Thus the duration time $\tD$ will    
be mainly determined by $\tsyn$ as shown in the left panel,    
especially for the light curves of 0.1, 1, and 10eV photons.    
Table 2 lists the duration properties   
of LC$_1$ and LC$_2$ for the non-expansion cases.    
   
{\it Relative Amplitude:} We find the light curves change with $\rsh$ in   
several factors.  First, for model A4, the condition of NE1 is satisfied;    
all the synchrotron timescales (for high energy light curves,   
see the right panel in Figure 3) are shorter than $\tltr$.    
The relative amplitudes are roughly approximated by equation (\ref{eq:amp-ne1}).    
With the decrease in the blob size, the relative amplitudes for high energy    
light curves roughly keep constant but they become very complicated for    
low energy light curves, because the condition NE1 is broken.     
For model A2, $\rsh=10^{15}$cm, $\tltr$ is shorter than any synchrotron   
timescales for lower energy light curves (left panel in Figure 3),    
then the condition NE3 is satisfied.  The relative amplitudes are given by   
equation (\ref{eq:amp-ne3}). The upper panel of the left side in Figure 3 indeed 
shows this property.  Increasing the size, the relative amplitudes   
become complicated as shown for models A3, A1, and A4.   
Second, the peak flux decreases with the size, if condition NE1 is satisfied.    
This property can be seen from the right panel of Figure 3.    
The left panels show the relative amplitude of peaks at different  
bands non-monotonously decreases with the size for low energy light curves. 
This reflects the changes of conditions in the blob due to the size of blob.  
Third, the duration is generally prolonged with the increase in the size.   
All the three features can be explained by the fact that the changes of   
the size of a blob modify the light travel time. When the size is large  
enough for a fixed frequency, the broaden light curves will be of small 
amplitude,  because the timescale of energy losses of electrons is long,   
which leads to the relative amplitudes of variations at lower frequency 
dominate over high frequency.   
   
{\it Time Lag:} The most prominent properties of these light curves are    
the time lags among different wavebands.  From Figure 3,    
we find $\tlag$ is very sensitive to the size with complicated behaviors.   
For the symmetric light curves there is zero-lag ($\tlag=0$),   
because the rising and falling timescales are given by $\tltr$   
as for the case of $\rsh=10^{17}$cm at 1, 2, 3, and 4keV wavebands.     
However for the asymmetric light curves, the time lag becomes complex. As 
we have already pointed out that the rising timescale is roughly determined    
by $\max (\tsyn, \tltr)$. Thus for high energy light curves the time lag 
$\tD$ is mainly affected by the energy losses, if the size is small enough 
as shown by the right panel of Figure 3, which does not show significant time lag,   
because the rising timescale is too short to display by a large scale time axis.   
It is evident to find that time lags increase significantly with the   
size for lower energy light curves from the left panel of Figure 3.   
We still find time lags for $\rsh=10^{15}$cm are too small to illustrate   
in Figure 3.  Since $\tsyn>\tltr$ for 0.1, 1, and 10eV photons,    
it is expected that time lag is mainly determined by the synchrotron energy losses.   
For intermediate cases $\tD$ will be jointly given by two processes, i.e.,   
light travel and synchrotron radiation.  Our numerical results confirm   
the qualitative analysis listed in Table 2.    
  
Here we have addressed the effects of the blob's size on the properties    
of light curves.  These features are key observational implications,    
as we have shown and listed above in detail. Because the properties of 
light curves are highly dependent on the observable frequencies, it is 
desired to compare multiwavelength light curves.

\subsubsection{Effects of magnetic fields}   
   
Equations (\ref{eq:amp-ne2}), (\ref{eq:amp-ne3}), and (\ref{eq:amp-ne4})    
already show the role of magnetic fields.     
The effects of magnetic field for the non-expansion cases are shown in    
Figure 4.  We change the magnetic field as a free parameter,    
while all other parameters are fixed.     
The role of magnetic field can be deduced from the timescale of synchrotron losses    
$\tsyn \propto B^{-3/2}$ [equation (\ref{eq:tausyn})].   
The stronger magnetic field, the shorter the timescale of synchrotron    
losses.  This leads to 1) the luminosity is enhanced by increasing   
magnetic field, 2) the rising timescale becomes shorter if the size   
is small enough.  We fix $\rsh=10^{16}$cm and compare the model A1   
with others to show the effects of magnetic field.   
Correspondingly, the peak flux reaches in advance by increasing the magnetic field.    
This leads to the shortening of time lag $\tlag$.    
This can be directly seen in Figure 4.  Of course the duration    
timescale $\tD$ also shortens by increasing magnetic field.   
If the magnetic field is suitable, symmetric light curves will appear by combined 
effects of synchrotron losses and light travel. For $\rsh=10^{16}$cm,   
we will obtain symmetric light curves, if we increase $B$ until $\tsyn<\tltr$.   
     
\subsubsection{Effects of electron energy index}   
   
Without a specific acceleration mechanism, we assume that the injection spectrum   
has a power-law with index $\alpha$. The equations in qualitative analysis 
indicate the importance of index  $\alpha$. The roles of the power-law index 
$\alpha$ of the injected electrons are shown in Figure 5.  From equation 
(\ref{eq:q0}) we know that the lower energy electrons are dominant over high 
energy electrons, if $\alpha>2$. The power radiated in lower energy will be 
larger than that in high energy bands. This is already shown in Figure 3, where 
we take $\alpha=2.5$ in several models. Thus the contribution to the rising 
side of light curves due to the synchrotron radiation of electrons shifted 
from higher energy can be neglected, because its number is much less than the 
initial lower energy electrons.  Time lag is very sensitive to the index $\alpha$.     
Lowering $\alpha$ leads to the postpone of the flux peak, and shortening time lag.   
The relative amplitude of peak fluxes decreases with the increase of $\alpha$.   
In such a case the relative amplitude of the peak   
fluxes would be related with the time lag.

\subsubsection{Effects of electron escape}   
   
The role of escape of electrons from the radiation region is considered. It is 
easy to find from equations (\ref{eq:el-dist}) and (\ref{eq:amp-ne4})    
that the number of electrons decreases with time.   
With the inclusion of the size effects, the escape have three factors:   
the vast decreases of fluxes, the shortening of time lag, and the asymmetry   
of light curves. These can be seen in Figure 6, where we fix the size   
of the blob as $\rsh=10^{16}$cm. The escape effect is small   
on high energy light curves, because $\tsyn<\tesc$.  For the cases with   
longer escape timescale, its effects can be neglected.  For example,   
$\tesc/\tltr=20$ is long enough. Its effects on time lag $\tD$ for low    
energy light curves is significant:  the decrease of $\tD$ with increasing    
$\tesc$ as well as the relative amplitude [see eq. (\ref{eq:amp-ne4})].

\subsubsection{Effects of injection timescale}   
   
The effects of injection timescale on the light curves can be deduced from   
equation (\ref{eq:qinj}).  The longer the injection, the lower   
amplitudes of peak fluxes as shown in Figure 7, because the total injected 
energy $E_0$ is fixed. The time that fluxes attain their peak will be 
postponed with increases of $t_2$. It can be found that $\tD$ decreases 
with increasing $t_2$. The symmetry strongly depends on the injection 
timescale $t_2$ as shown by right-upper panel. The reason is simple: three 
timescales, $\tltr$, $\tsyn$, and $t_2$, are competing in the blob.    
If $t_2$ is short enough compared with $\tsyn$ and $\tltr$,    
the injection can be regarded as impulsive injection.    
Then the light curves will be determined by the other two timescales $\tsyn$ and    
$\tltr$.  For longer duration of injection, the light curves will be complicated.   
For a case of $t_2 > \max(\tsyn, \tltr)$, the rising timescale will be determined   
by the injection.  Generally speaking, the rising timescale will be   
given by $\max(t_2, \tsyn, \tltr)$, whereas the falling timescale will   
be given by $\max(\tsyn, \tltr)$ as we already discussed.   
      
\subsubsection{Evolution of photon spectrum}   
   
Figure 8 shows the plot of photon index ($\aph$) vs. luminosity    
for model A1.  The evolution direction is clockwise.    
We find the behaviors of the evolution at 1, 2, 3, and 4keV are quite similar.    
The photon index increases with luminosity (phase 1) before it reaches a maximum,    
and then the index stays constant (phase 2).  Phase 1 is caused by injection   
and energy loss; the decrease of $\alpha$ in phase 1 occurs, because high energy   
electrons lose energy quickly and a break appears in the electron spectrum   
as shown in Figure 2 (left panel), while the increase in the photon energy flux   
is due the injection of electrons.   
On the other hand, in phase 2, the shape of the electron spectrum is steady   
but the total electron number decreases, which leads to the decrease in the photon    
energy flux. It can be found that $\Delta \aph\approx 0.2$ at 1keV, and   
$\Delta \aph$ decreases with increasing photon energy.

\subsection{Expanding Blob}   
   
We have pointed out that the non-expanding model is in fact a homogeneous   
model, because the geometry and magnetic field do not change during   
the evolution of a flare.  When a blob expands, although the expansion 
occurs at a constant speed and the density is homogeneous everywhere 
inside the shell in our model, the observer will see an inhomogeneous blob,   
considering the photon propagation; the received photons from a blob come 
from different parts of the blob and the different travel time from different 
parts results in emission from an effectively inhomogeneous blob.   
Then the light curves of the expanding blob show very complicated behaviors.   
We refer the expanding blob as modified homogeneous model.     
All three timescales $\tltr$, $\tsyn$, and $\tesc$ change with time.    
These result in complicated situations for the expanding blob.   
  
The expansion due to the internal energy can be described by Sedov's   
self-similar solution; the radius of the shocked shell reads   
\begin{equation}   
\rsh=\rsh^0\max\left[1,\left(\frac{t}{t_1}\right)^{2/5}\right] \, .   
\end{equation}   
The parameter $t_1$ is assigned for the validity of the self-similar solution.   
The exact value of this parameter should be obtained from solving   
the dynamics of a blob. Here we simply take it as an unimportant parameter    
in our model.  But we constrain it by non-relativistic expansion.   
This gives   
\begin{equation}   
t_1\geq \frac{2}{5}\frac{\rsh^0}{\beta_0 c}\approx \frac{\rsh^0}{c} \, ,   
\end{equation}   
where $\beta_0 c$ is the expansion velocity at $t_1$.    
We will carry out the research for the expansion with    
relativistic speed in future.   
We take the index $n=1, 2$ in equation (\ref{eq:mag-exp}).   
The initial speed is set by $v_0=\rsh^0/t_1$.   
   
We calculate the models listed in Table 4 in order to find the properties of   
an expanding blob.   
   
\subsubsection{Effects of magnetic field index}   
   
The index $n$ of magnetic field will strongly affect the properties of light   
curves.  We can understand these effects from equations    
(\ref{eq:syn-time}) -- (\ref{eq:nuc}). Especially for $n=2$, when expansion 
$\rsh=2\rsh^0$, synchrotron radiation timescale $\tsyn$ and its    
mono-electron power $\dot{\gamma}$ will be reduced to 1/16 of the non-expansion 
case. It is expected that relative amplitude, time lag, symmetry, and duration 
time are distorted seriously. Thus the effects of expansion will be of significantly   
observational implications. Figure 9 shows the light curves of models exp-A1 and 
exp-C1, showing the dependences of light curves on the magnetic field index $n$.  
In the left panel of Figure 9, it is seen that the shapes of light curves are very 
sensitive to the index $n$. The duration $\tD$ for $n=1$ is fundamentally different 
from that for $n=2$. Comparing exp-C1 and exp-A1 with non-expanding A1 models    
(the parameters are the same besides expansion), we can find interesting 
differences. Between A1 and exp-C1, we find four properties as follows.  
1) the luminosity is significantly lowered by expansion,    
because the magnetic field weakens via expansion as $B\propto r^{-n}$.   
2) the relative amplitudes are changed,   
$L^{\rm pk}$(100eV)/$L^{\rm pk}{\rm (10eV)} > 1$ for expansion,    
whereas $L^{\rm pk}$(100eV)/$L^{\rm pk}{\rm (10eV)} < 1$ for   
non-expansion case.  The similar properties can be found by comparing    
$L^{\rm pk}$(100eV) and $L^{\rm pk}$(1eV) etc. 3) Duration time $\tD$ is 
changed significantly.  Model exp-C1 has much longer $\tD$ than A1, because 
the synchrotron energy loss rate become smaller for weakening magnetic field.    
4) Time lag $\tlag$ is shortened via expansion, because the flux peak time occurs 
in advance.  This effect is complicated, because the rising timescale will be 
controlled by three timescales such as $\tsyn$, $\tltr$, and $\texp$.    
For high energy light curves, $\tsyn$ will be   
shorter than $\tltr$ and $\texp$.  Thus the rising timescale will be   
mainly determined by $\tsyn$ as comparing with non-expansion cases.    
Also time lag $\tlag$ is expected [it is not significant for keV light curves, 
but we can compare $L^{\rm pk}$(0.1keV) and $L^{\rm pk}$(1keV)].   
It should be noted that $t_1$ has a strong effect on the peak flux time.  
But it is beyond the scope of this paper to discuss the value of $t_1$,  
because we do not solve hydrodynamics of the expansion.  
  
Between A1 and exp-A1, light curves are almost changed completely by the   
expansion with $n=2$.  First, output luminosity weakens. 2) The relative    
amplitudes are changed similar to exp-C1. 3) Duration time $\tD$ is completely    
different between A1 and exp-C1.  The luminosity weakening is due to the   
the rapidly weakening magnetic field rather than the energy losses of    
electrons. Thus the duration will be controlled by expansion. 
The falling profiles are $t^{-m}(m\approx 0.6\sim 0.7)$ rather than linear 
decrease, namely they become dim with powerlaw of time.  From 
equation (\ref{eq:syn-time}), we know the synchrotron timescale becomes very 
long ($\tsyn\propto \rsh^{4}$). This leads to the decrease in luminosity owing 
to the decrease of magnetic field rather than the energy loss of electrons.    
Thus this is completely different from the other   
cases for $n=1$.  Under such a circumstance, we quantitatively estimate   
the light curves from equation (\ref{eq:e-loss12});     
$\nu F_{\nu}\propto \DE/\texp$, then   
\begin{equation}   
\nu F_{\nu}\propto \nu^{(1-\alpha)/2}t^{(1-\alpha)n/5} \, ,   
\end{equation}   
where we use $\nu\propto B\gamma^2 \propto t^{-2n/5}\gamma^2$.   
We see the falling profile  $t^{-0.6}$ (for $n=2$ and   
$\alpha=2.5$) and the frequency-dependence   
$\nu^{(1-\alpha)/2}$.  This crude approximation is in agreement with  
the detail calculations shown in Figures 9 and 10 for the model exp-A1,   
exp-C1, and exp-C2. 4) There is almost no significant time lag between   
the light curves both for high and low energy light curves. This is   
because $\texp\ll \tsyn$. 5) The peaks are very sharp. This is a new    
characteristics caused by index $n$. Expansion leads to strong  
$n$-dependence of synchrotron timescale. After $t = t_1$, the decline  
of light curves is caused by weakening magnetic field rather than energy  
losses of relativistic electrons.   
   
\subsubsection{Effects of initial expansion speed}   
   
We fix the initial radius of a blob $\rsh^0$, but adjust $t_1$ in order  
to test the effects of initial speed.  Figure 11 shows the light curves   
for different velocity $\beta_0 = 0.22 c$ ($t_1 = 6 \times 10^5$ s)   
and $\beta_0 = 0.17 c$ ($t_1 = 8 \times 10^5$ s).    
   
Generally the expansion enlarges the size of a blob, and thus leads to  
similarity with the cases of larger radius if the expansion is fast enough.  
Figure 10 clearly shows the effects of initial expansion speed. The effects  
are clearer for high energy light curves (keV bands).  The lower initial  
velocity $\beta_0$, the faster the peak flux is attained. The peak flux  
significantly lowers, because the faster expansion leads to rapid decrease  
in magnetic field. Then the duration time $\tD$ for faster initial expansion  
speed is shorter  than that for slower initial expansion speed.     
On the other hand, time lag is not changed much by initial expansion speed.    
Time lag $\tlag$ is jointly controlled by $\tsyn$, $\tltr$, and $\texp$.     
We summarize in Table 2.  For high energy light curves   
the rising timescale may be mainly determined by $\tsyn$, and the falling   
timescale is controlled by $\tltr$ and $\texp$.  Thus the symmetry of   
light curves will be generally broken. The expanding blob will always   
have asymmetric light curves, because the timescales are changing during  
the expansion.  
     
\subsection{Evolution of spectrum}   
   
Figure 11 illustrates the trajectory in the photon index vs. flux plane for   
an expanding blob.  The upper panel shows the evolution of the spectrum for $n=2$.    
It can be seen that the behaviors are quite different from those of   
the non-expansion cases.   
The spectral evolution is similar for non-expanding and expanding blobs,   
before the luminosity attains its maximum.  The photon index $\aph$   
increases with increasing luminosity. However, the evolution behaviors   
are different after then.  We see $\aph$ continues to increase   
with decrease in luminosity for an expansion blob.    
The lower panel shows the comparison for the cases of $n=1$ and $n=2$.     
We find that $\aph$ becomes larger values for $n=2$.  We also find that the behavior   
difference becomes large with decreases of photon energy.    
These results are mainly because of the decrease in magnetic field.

\section{Conclusions}   
   
We presented a very simple Sedov-expansion model of a coasting blob and studied   
the received spectrum radiated from it.  The multiwaveband light curves  
were calculated for the lower energies such as 0.1, 1, 10, and 100 eV,  
and higher energies such as 1, 2, 3, and 4 keV. 
Main four processes determine the observed light curves, namely, 
the evolution of the energy spectrum of relativistic electrons,    
the size (initial and enlarging size) of the radiation region,  
bulk motion of the blob, and electron escape from the radiation region.   
According to the timescales, $\tsyn$, $\tltr$, $\texp$, and $\texp$,  
we find 11 kinds of light curves, which are listed in Table 2 with  
a concise summary of their properties including symmetry, duration, relative  
amplitude, and time lag.  
To facilitate to compare our model with observations, we stress the followings: 
 
(1) In X-ray band the rising profile of light curves is determined by   
the timescale of light travel in a blob, and the declining profile is symmetric,   
if the timescale of synchrotron radiation is short.   
   
(2) In EUV band the rising timescale of the light curve is the sum of   
$\tau_{\rm ltr}$, $\tau_{\rm syn}^{\rm EUV}$,   
and the lifetime of electrons radiating EUV photons.    
The declining profile is determined by $\tau_{\rm syn}^{\rm EUV}$.   
   
(3) The observed UV light curve profile is determined by the whole processes,   
and the rising timescale is roughly the sum of $\tau_{\rm ltr}$,   
$\tau_{\rm syn}^{\rm EUV}$, and $\tau_{\rm syn}^{\rm UV}$.    
The declining timescale is determined by the minimum of $\tau_{\rm esc}$   
and $\tau_{\rm syn}^{\rm UV}$.   
   
(4) Light curves from an expanding blob depend on how magnetic field decreases   
with expansion.  When the decrease of luminosity is steep, it may be because of   
the rapid decrease of magnetic field and it is possible   
to estimate the index $n$ of magnetic field.   
 
(5) The relationship between the duration time, rising time, decline time,   
and the relative amplitudes in different wavebands is expected to    
be found from the multiwaveband observations.  It is useful to test the undergoing   
flare mechanisms by using the model presented here.   
 
(6) For the rapid decays of magnetic field, the falling profiles can be 
described as $\nu F_{\nu}\propto \nu^{(1-\alpha)/2}t^{(1-\alpha)n/5}$, 
which is significantly different from the slowly decaying one without expansion.  
Thus the profile can be used to test the magnetic field in the blob. 
   
These properties apply for the emission spectrum which has the synchrotron tail   
in X-ray bands; also note that this situation depends on the value of    
$\gamma_{\rm max}$ and $\cal D$, too.   
   
In this paper we calculated the light curves of an expanding blob interacting with   
its surroundings. The electron injection function was obtained analytically    
and self-consistently based on the Sedov's self-similar expansion.    
We neglected the re-acceleration of electrons which lose energy    
via synchrotron radiation.  Since the energy loss timescale may be comparable    
with the expansion timescale (or the injection timescale),    
the acceleration region and the radiation region in fact are almost    
in the same region.  Then the re-acceleration of electrons is possible    
(Schlickeiser 1984).  The pile-up energy will shift with time.     
We will study this subject in future.   
The time-dependent injection with a temporal profile may lead to more complex   
light curves. It is needed to study the detailed acceleration process to interpret   
the time-dependent injection pattern, which is far beyond the scope of   
the present paper.   
In our present calculations we used the same approximation made by Dermer    
(1998) and Georganopoulos and Marscher (1998) who neglected inverse   
Compton scattering. The validity of this approximation has been discussed   
in Georganopoulos and Marscher (1998).  If the expansion is relativistic   
then the difference of Doppler factor at different locations in a blob can not   
be neglected.  We will study these in future, too.

\acknowledgements {J.-M. W. is supported by the ``Hundred Talents Program of CAS.''  
This research is financed by the Special Funds for Major State Basic Research    
Projects and NSF of China. He thanks the useful discussions with Y.Y. Zhou, M. Wu, 
T.P. Li, J.L. Qu, and H. Netzer.}

\newpage   
   
\vskip 1cm   
   
\begin{center}   
{Table 1. The input parameters in the present model}   
\end{center}   
\begin{center}   
\begin{tabular}{rlcl}\hline \hline   
1.&$B_0$             &......&blob's initial magnetic field\\   
2.&${\cal D}$        &......&blob's Doppler factor\\   
3.&$E_0$             &......&Energy of relativistic electrons injected into a blob\\   
4.&$\alpha$    &......&index of relativistic electrons at injection time\\   
5.&$\gamma_{\rm max}$&......& maximum energy of electrons in a blob\\   
6.&$\gamma_{\rm min}$&......& minimum energy of electrons in a blob\\    
7.&$r_{\rm sh}^0$    &......& initial radius of shocked shell of a blob\\     
8.&$\tesc$       &......& timescale of electron escape from the radiation region\\    
9.&$t_1$    &......& timescale of expansion before self-similar solution is applied\\    
10.&$t_2$           &......& timescale of injection\\    
11.&$\beta_0$       &......& initial expansion velocity of a blob in units of $c$\\    
12.&$n$             &......& index of magnetic filed\\ \hline   
\end{tabular}   
\end{center}

\clearpage    
\begin{figure}   
\vspace{3.5cm}   
\epsscale{0.1}      
\plotfiddle{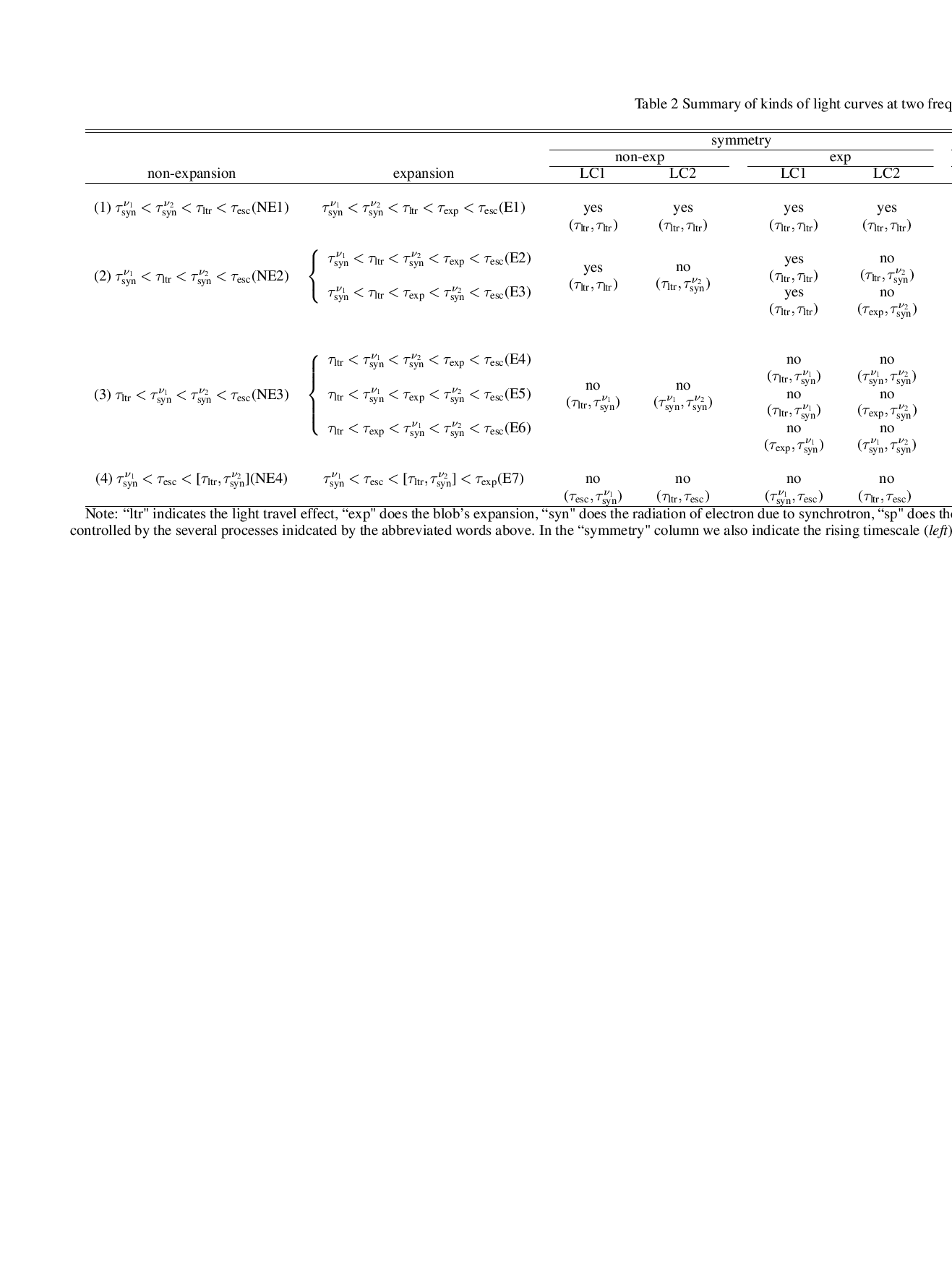}{190pt}{-90}{70}{70}{-450}{530}     
\end{figure}    
   
\clearpage   
\bigskip   
   
\def\rshsix{$r_{\rm sh}^0(\times 10^{16})$}   
\def\tetd{$\tesc/\tau_{\rm ltr}$}   
\def\ttwo{$t_2(\times 10^6)$}   
\def\tone{$t_1(\times 10^5)$}   
   
\begin{center}   
Table 3. Parameter values for non-expansion models   
\end{center}   
\begin{center}   
\begin{tabular}{lllllllllllllllll}\hline \hline   
Model                 &A1 &A2 &A3 &A4 &B1 &B2 &B3 &B4 &C1 &C2 &D1 &D2 &E1 &E2\\ \hline   
$\alpha$              &2.5&~  &~  &~  &~  &~  &~  &~  &2.5&~  &1.5&2  &2.5&~ \\   
\ttwo $\,$ s          &5  &~  &~  &~  &~  &~  &~  &~  &5  &~  &~  &~  &2.5&1 \\   
$B_0 \,{\rm G}$       &0.1&~  &~  &~  &0.2&0.3&0.4&0.6&0.1&~  &~  &~  &~  &~ \\   
\rshsix ${\, \rm cm}$ &1  &0.1&5  &10 &1  &~  &~  &~  &1  &~  &~  &~  &~  &~ \\   
\tetd                 &10 &~  &~  &~  &~  &~  &~  &~  &5  &20 &10 &~  &~  &~ \\ \hline    
\end{tabular}   
   
Note: The default values of the parameters are referred to the same   
as their left values in this table.   
\end{center}

\bigskip\bigskip\bigskip   
   
\begin{center}   
Table 4. Parameters for expansion models   
\end{center}   
\begin{center}   
\begin{tabular}{lllllll}\hline \hline   
Model           &exp-A1  &exp-A2  &exp-C1  &exp-C2\\ \hline   
$\alpha$        &2.5     &~       &~       &~ \\   
\ttwo   $\,$ s  &5       &~       &~       &~ \\   
$B_0$ G         &0.1     &~       &~       &~ \\   
$n$             &2       &~       &1~      &~ \\   
\rshsix $\,$ cm &1       &~       &1  ~    &~ \\   
\tetd           &10~     &~       &~       &~ \\    
\tone $\,$ s    &6~      &3~      &6       &8~\\ \hline   
\end{tabular}   
   
Note: The default values of the parameters are referred to the same   
as their left values in this table.   
\end{center}

\newpage 
\begin{figure}   
\vspace{3.5cm}   
\epsscale{0.7}      
\plotfiddle{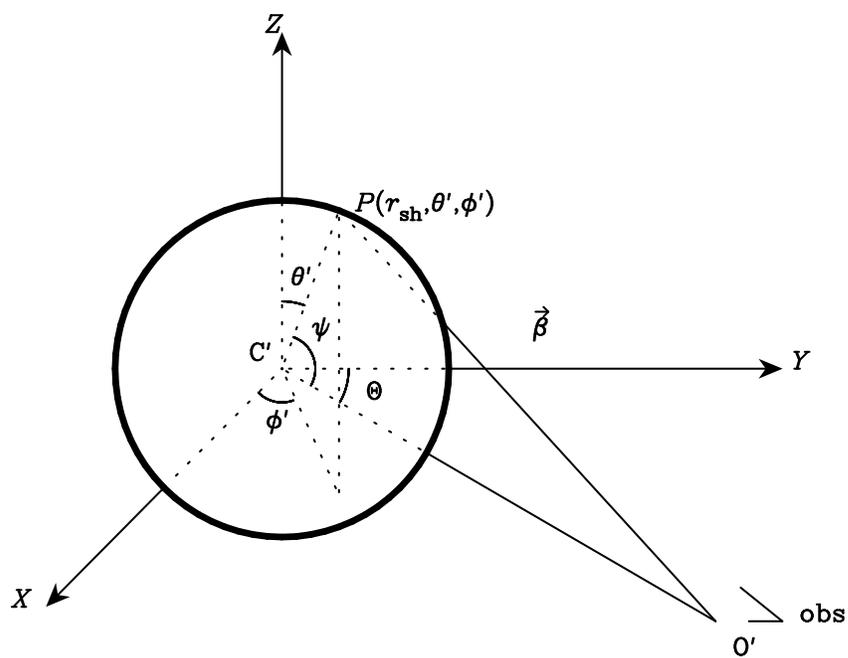}{190pt}{0}{70}{70}{-190}{-120}   
\vspace{20mm}   
\caption{   
The illustration of the expanding blob in its coasting frame.    
The black shell is the shock formed via the collision between   
the expanding blob and the inter-clouds media in broad line region.    
The shocked shell is very thin compared with the radius of the blob.   
The photons arrive at the observer at different time due to the different  
distances.   
Here $\Theta$ is the angle between the line of sight and the blob motion.}  
\label{fig1}   
\end{figure}    
   
\newpage 
\begin{figure}   
\vspace{3.5cm}   
\epsscale{0.5}      
\plotfiddle{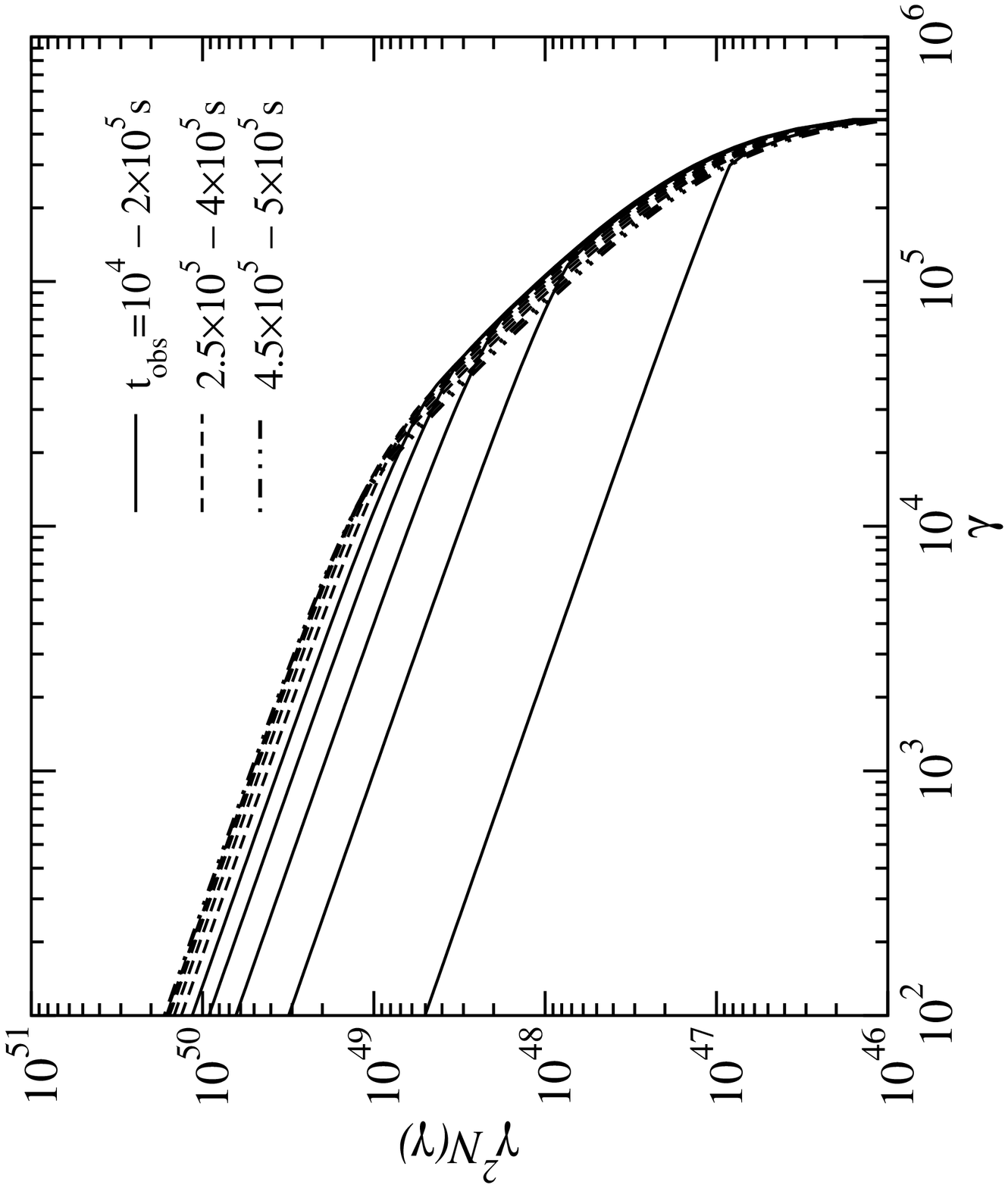}{190pt}{-90}{37}{40}{-250}{160}   
\plotfiddle{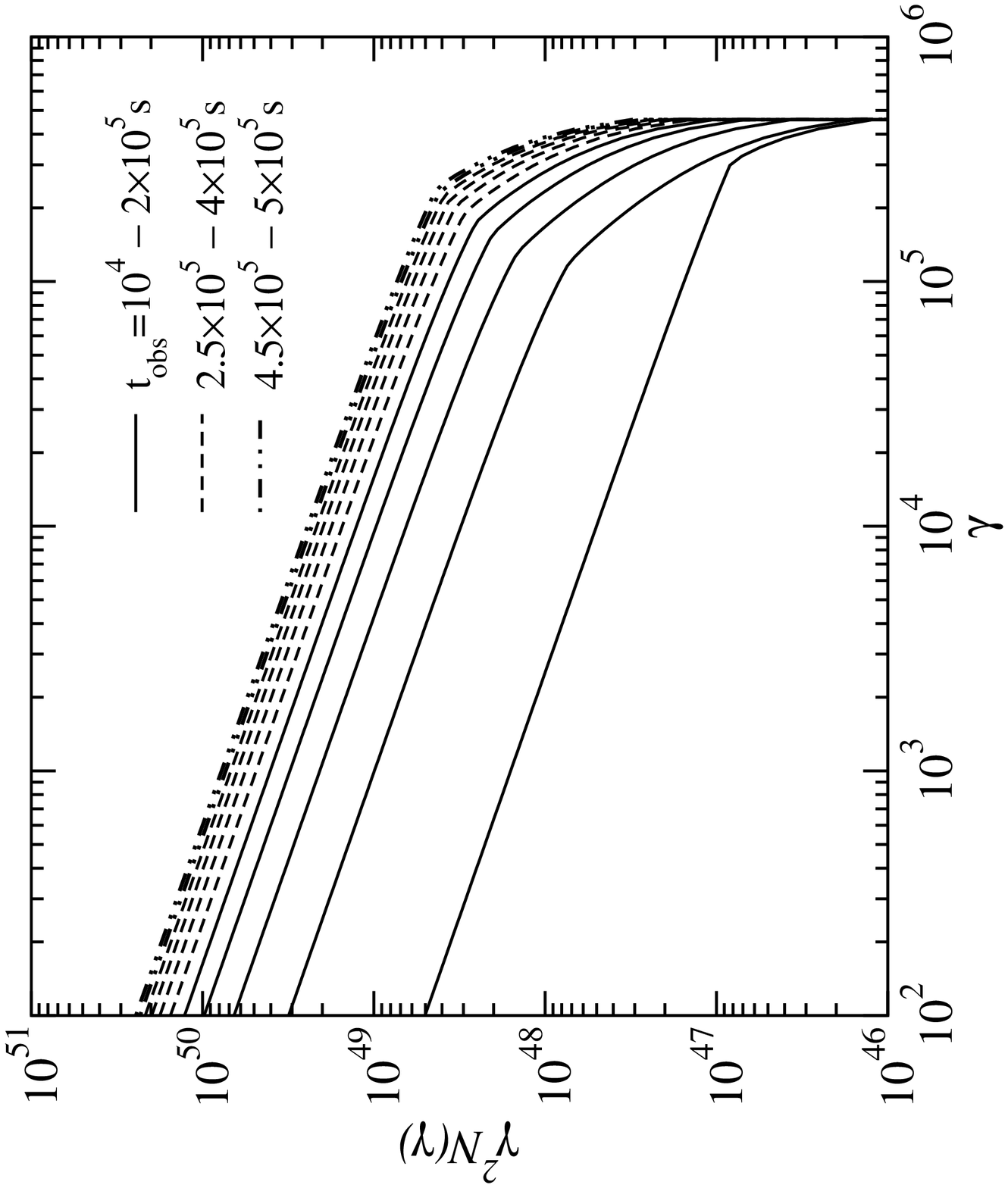}{190pt}{-90}{37}{40}{-20}{360}   
\vspace{1cm}   
\caption{The evolution of the electron energy spectrum. Left panel (Fig. 2a)  
is for a case of non-expansion (model A1), whereas the right panel (Fig. 2b) is for  
an expansion case (model exp-A1).   
For model A1, a broken power law appears after a certain time interval   
which is determined by timescales of energy loss and escape.  
For model exp-A1, the break energy is higher than that of model A1,  
because the energy loss is slower owing to the decreasing magnetic field.}   
\label{fig2}   
\end{figure}    
   
\newpage 
\begin{figure}   
\vspace{3.5cm}   
\epsscale{0.5}    
\plotfiddle{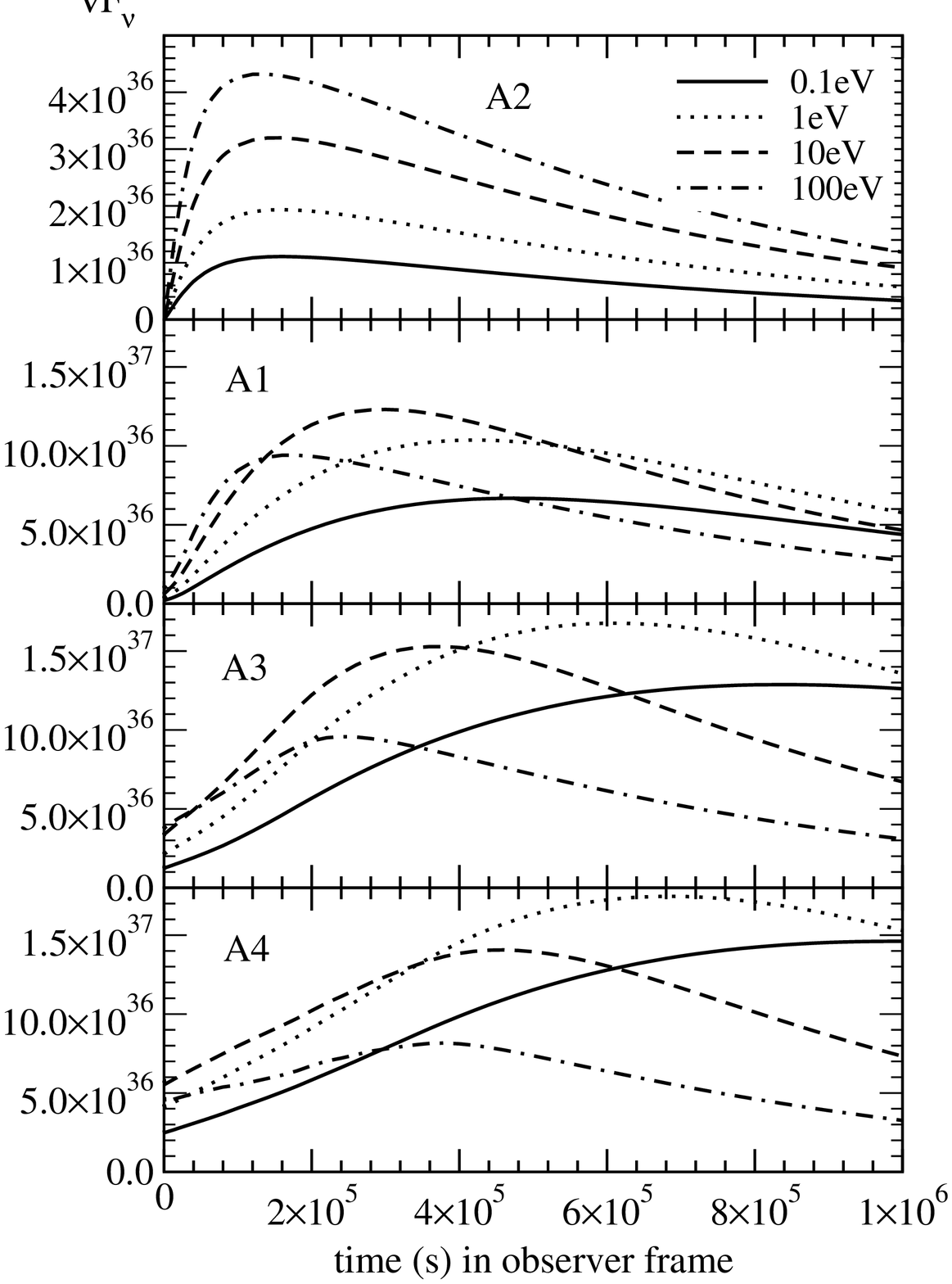}{190pt}{0}{50}{50}{-275}{-100}   
\plotfiddle{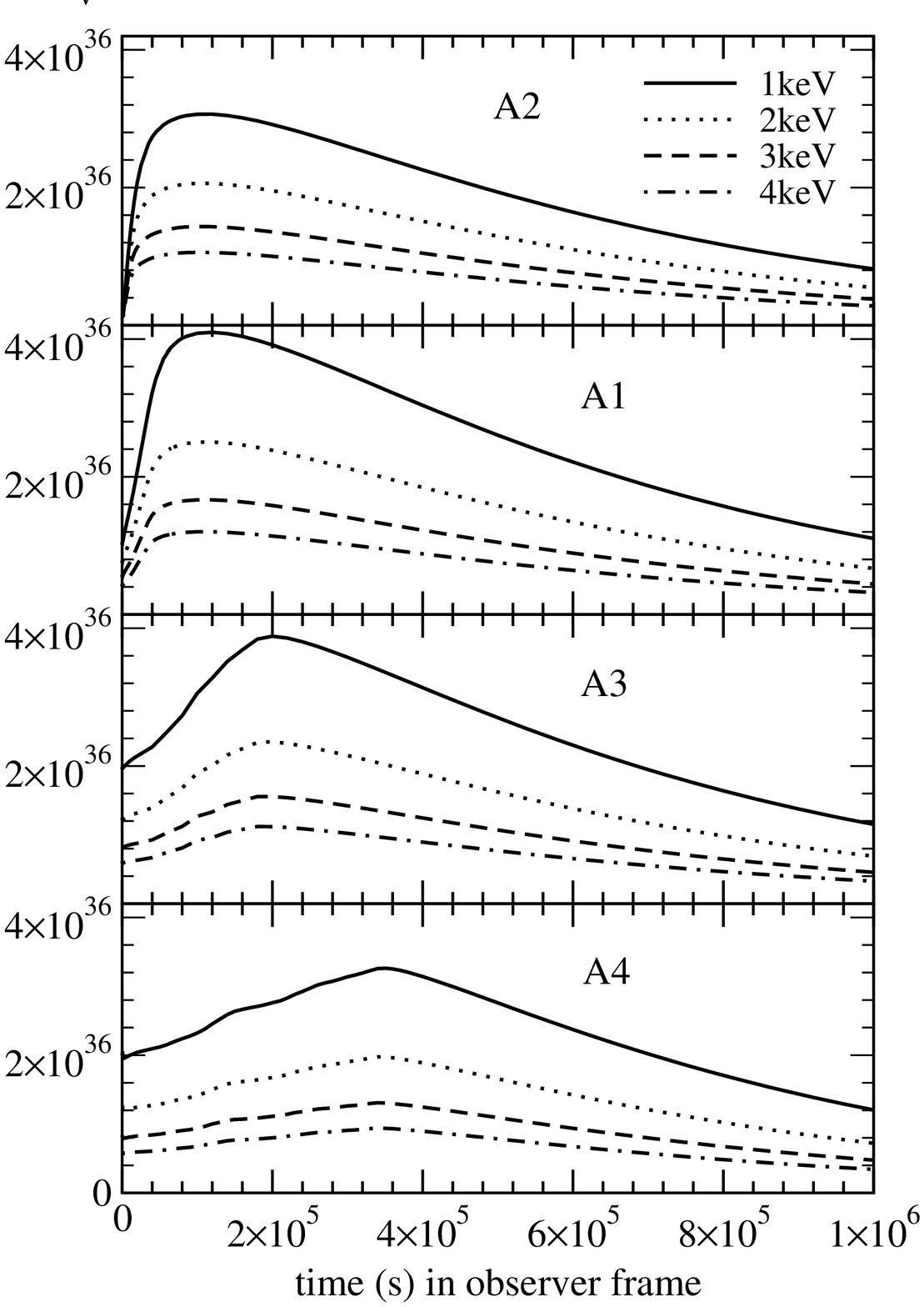}{190pt}{0}{50}{50}{-5}{100}   
\vspace{-5.1cm}   
\caption{The light curves for non-expansion blob for different values of  
$r_{\rm sh}$; $r_{\rm sh} = 10^{15}$, $10^{16}$, $5 \times 10^{16}$,  
and $10^{17}$ cm from top to bottom.  
The values of parameters for the models are listed in Table 3.   
}   
\label{fig3}   
\end{figure}    
   
\newpage 
\begin{figure}   
\vspace{3.5cm}   
\epsscale{0.5}   
\plotfiddle{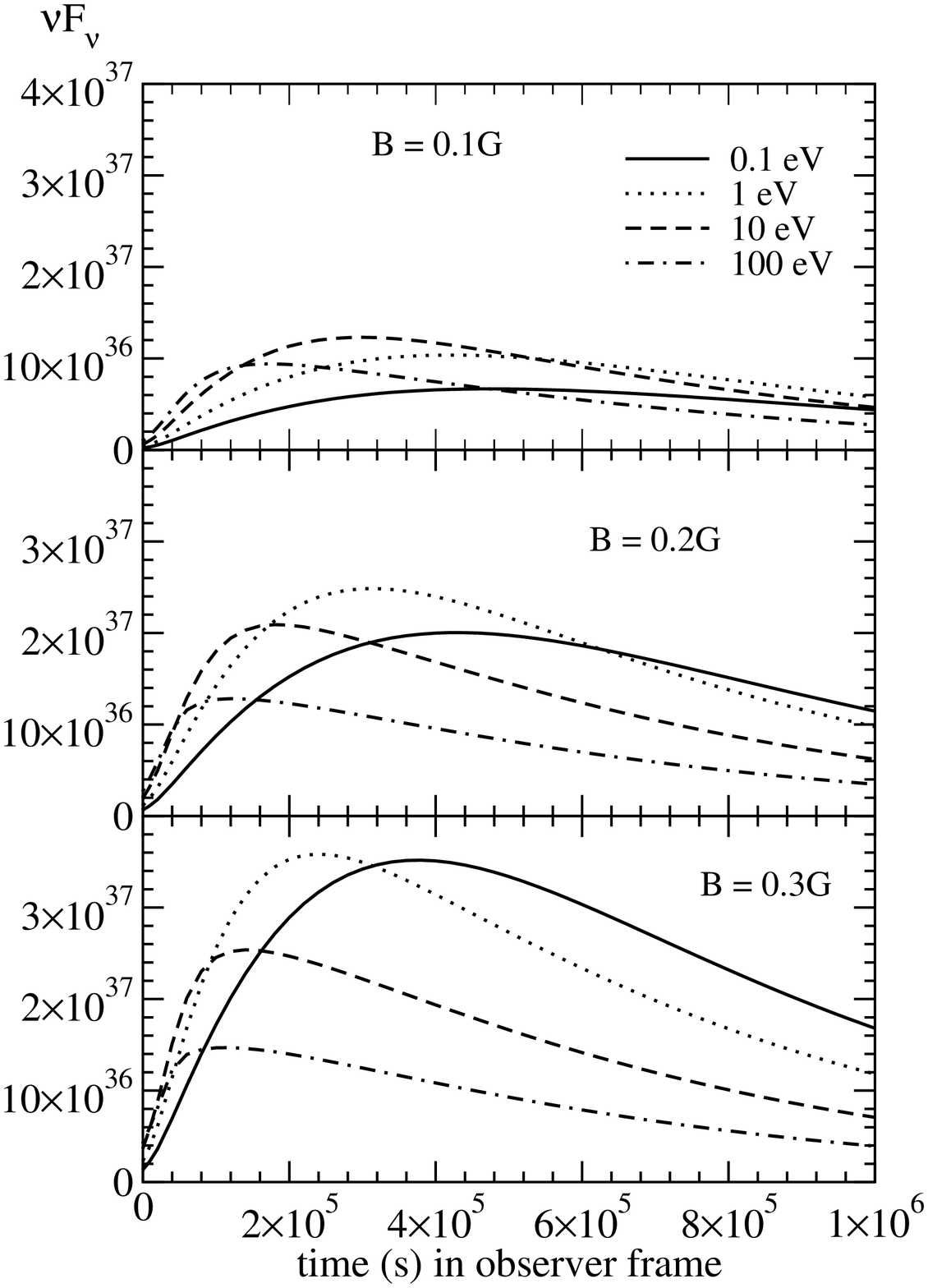}{190pt}{0}{45}{45}{-250}{-100}   
\plotfiddle{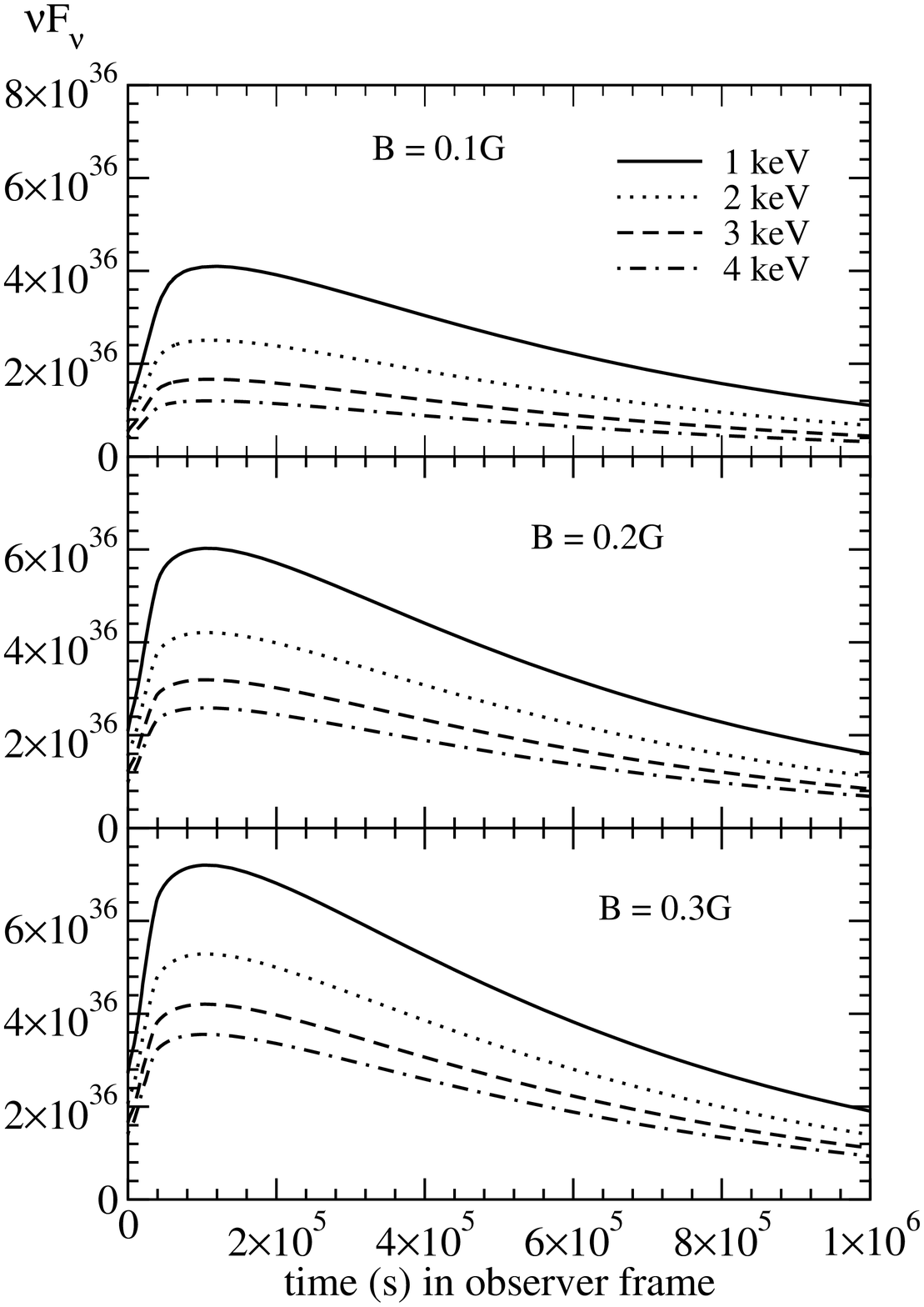}{190pt}{0}{45}{45}{10}{100}   
\vspace{0.1cm}   
\caption{The light curves for different magnetic field for a non-expanding  
blob: model A1, B1, and B2 from top to bottom.}  
\label{fig4}   
\end{figure}    
   
\newpage 
\begin{figure}    
\vspace{3.5cm}   
\epsscale{0.5}   
\plotfiddle{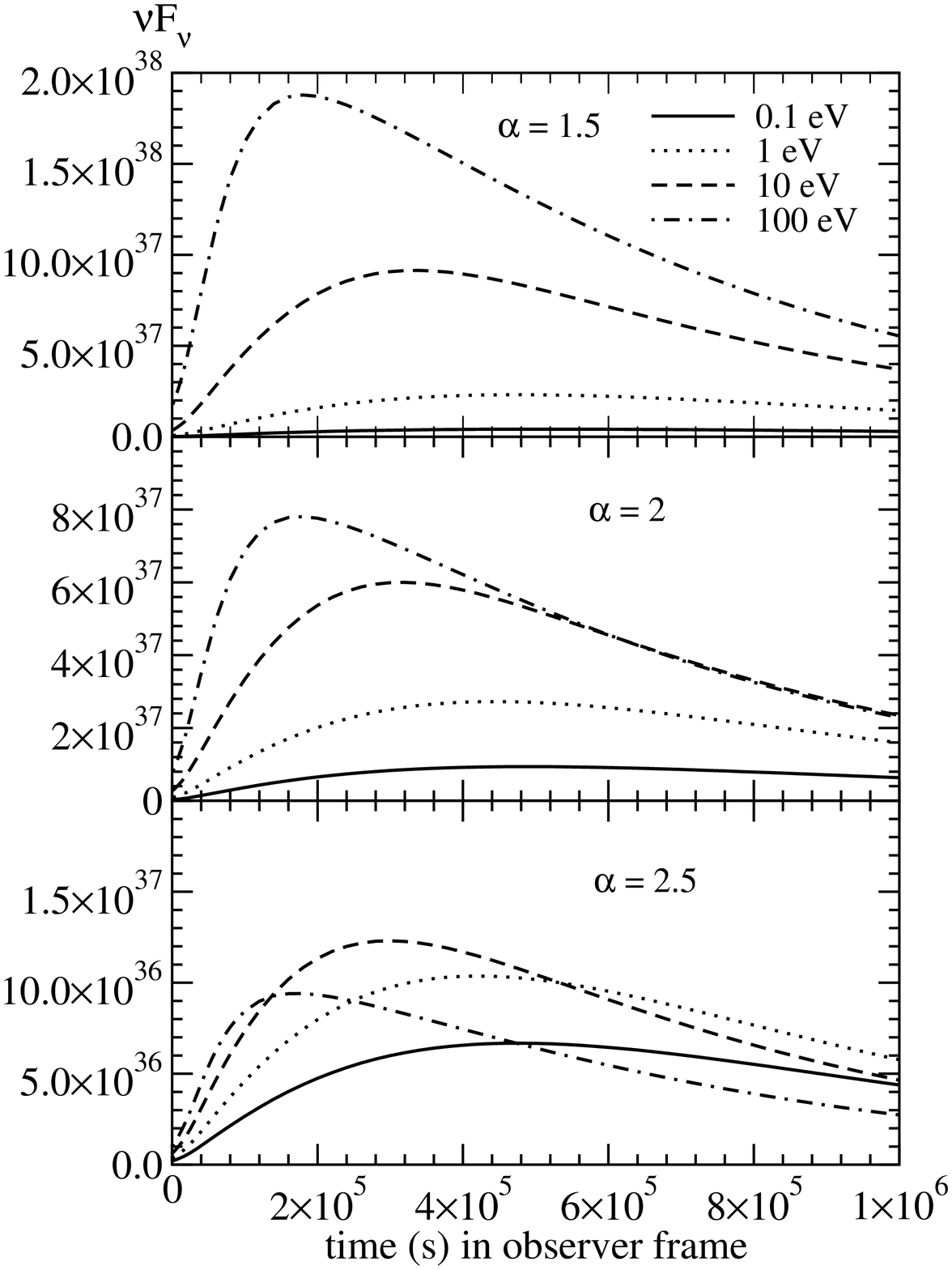}{190pt}{0}{45}{45}{-250}{-100}   
\plotfiddle{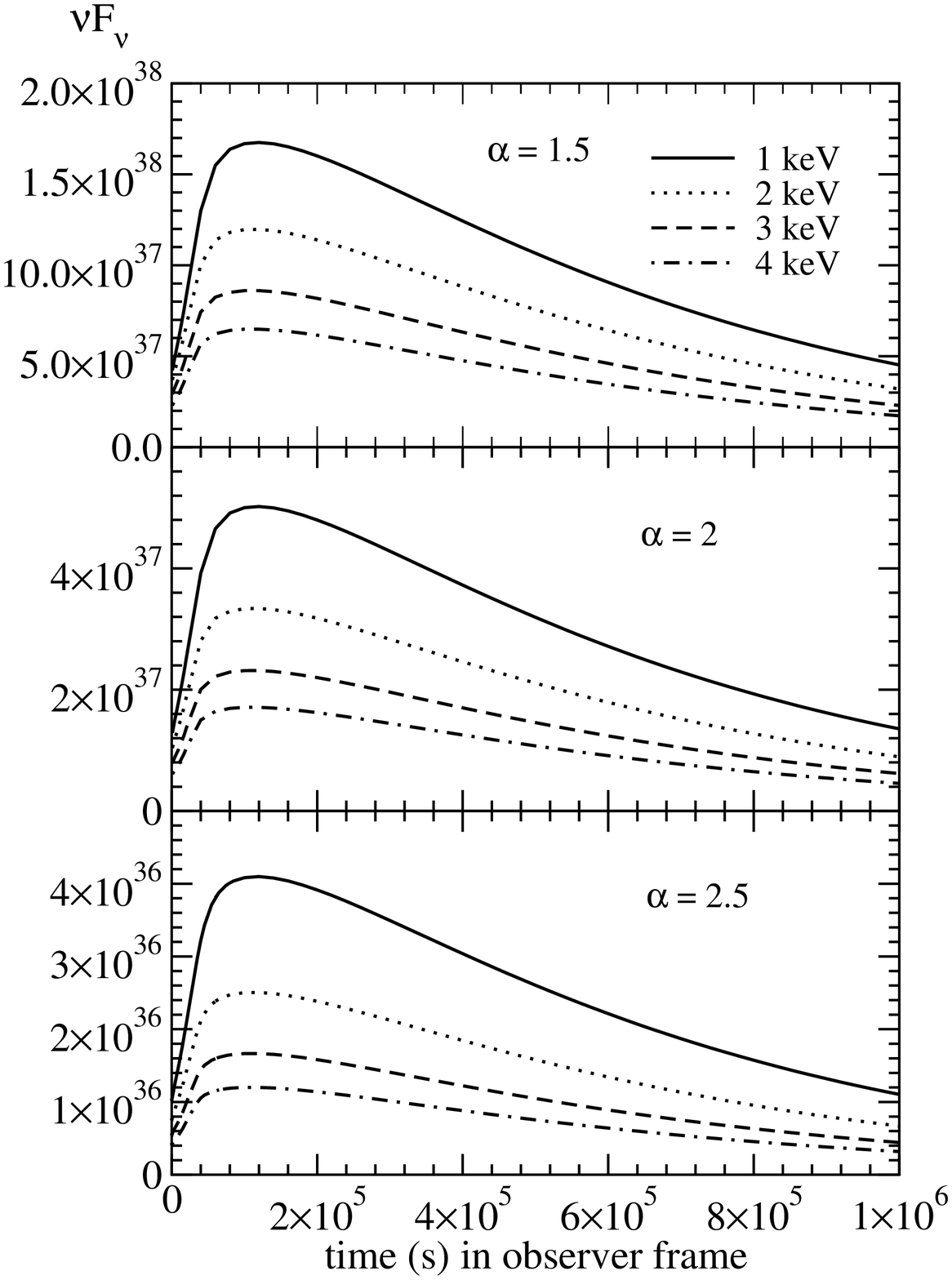}{190pt}{0}{45}{45}{10}{100}   
\vspace{0.1cm}   
\caption{The light curves for different values of the index of the injected electron  
spectrum for a non-expanding blob: model D1, D2, and A1 from top to bottom.}  
\label{fig5}   
\end{figure}    
   
\newpage 
\begin{figure}   
\vspace{3.5cm}   
\epsscale{0.5}    
\plotfiddle{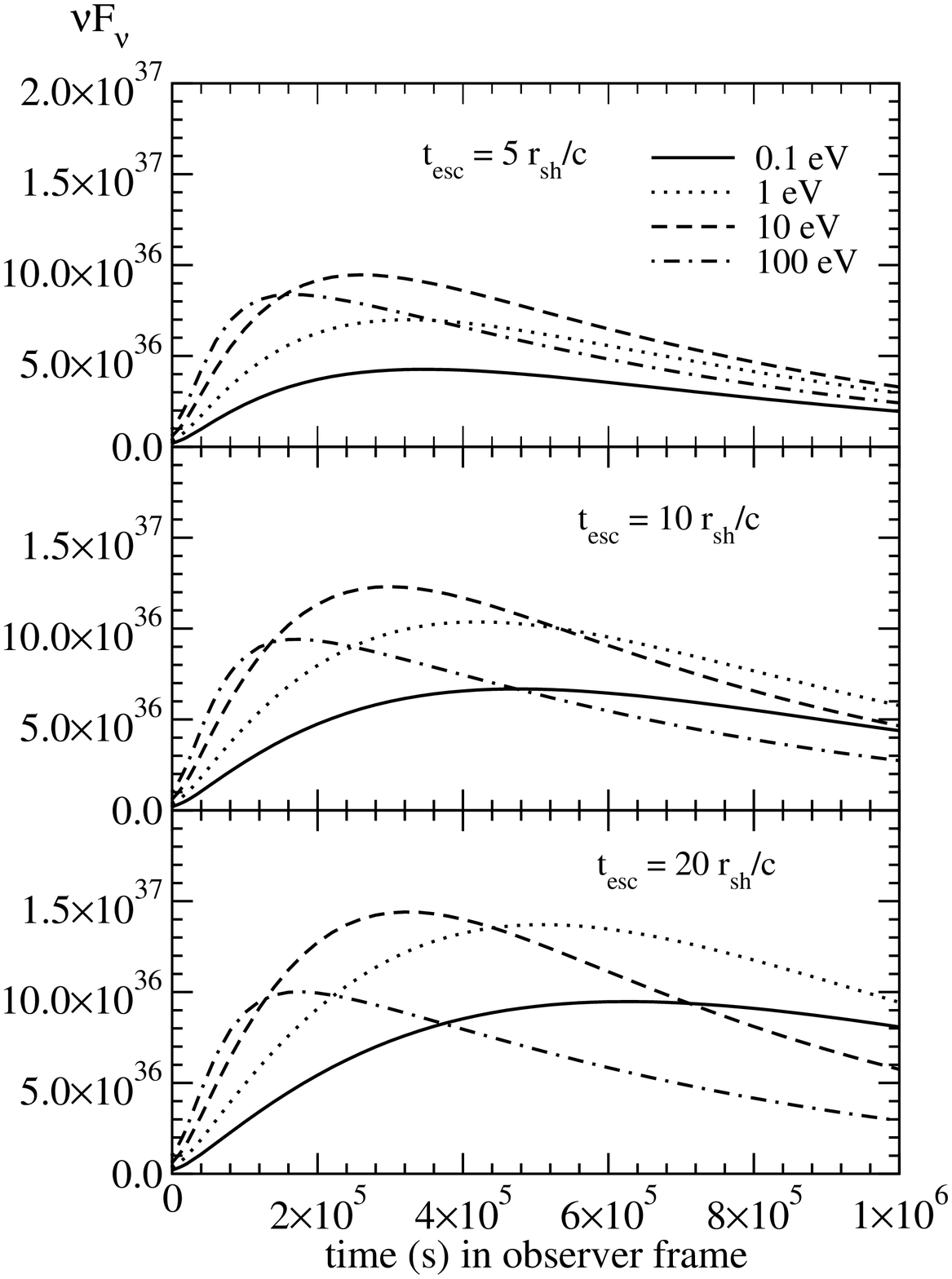}{190pt}{0}{45}{45}{-250}{-100}   
\plotfiddle{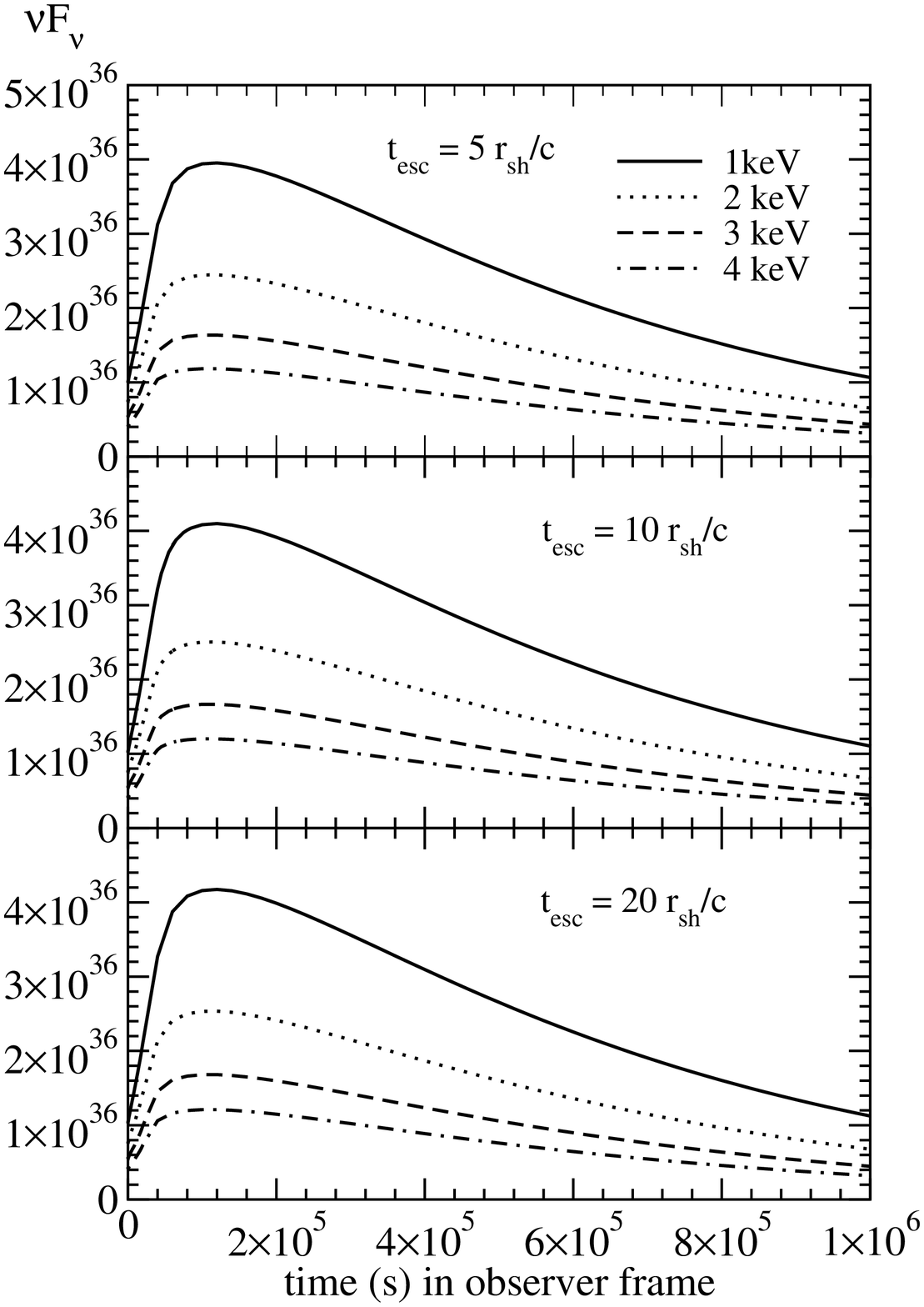}{190pt}{0}{45}{45}{10}{100}   
\vspace{0.1cm}   
\caption{The light curves for different escape timescales for   
a non-expanding blob: model C1, A1, and C2 from top to bottom.}   
\label{fig6}   
\end{figure}    
   
\newpage 
\begin{figure}   
\vspace{3.5cm}   
\epsscale{0.5}    
\plotfiddle{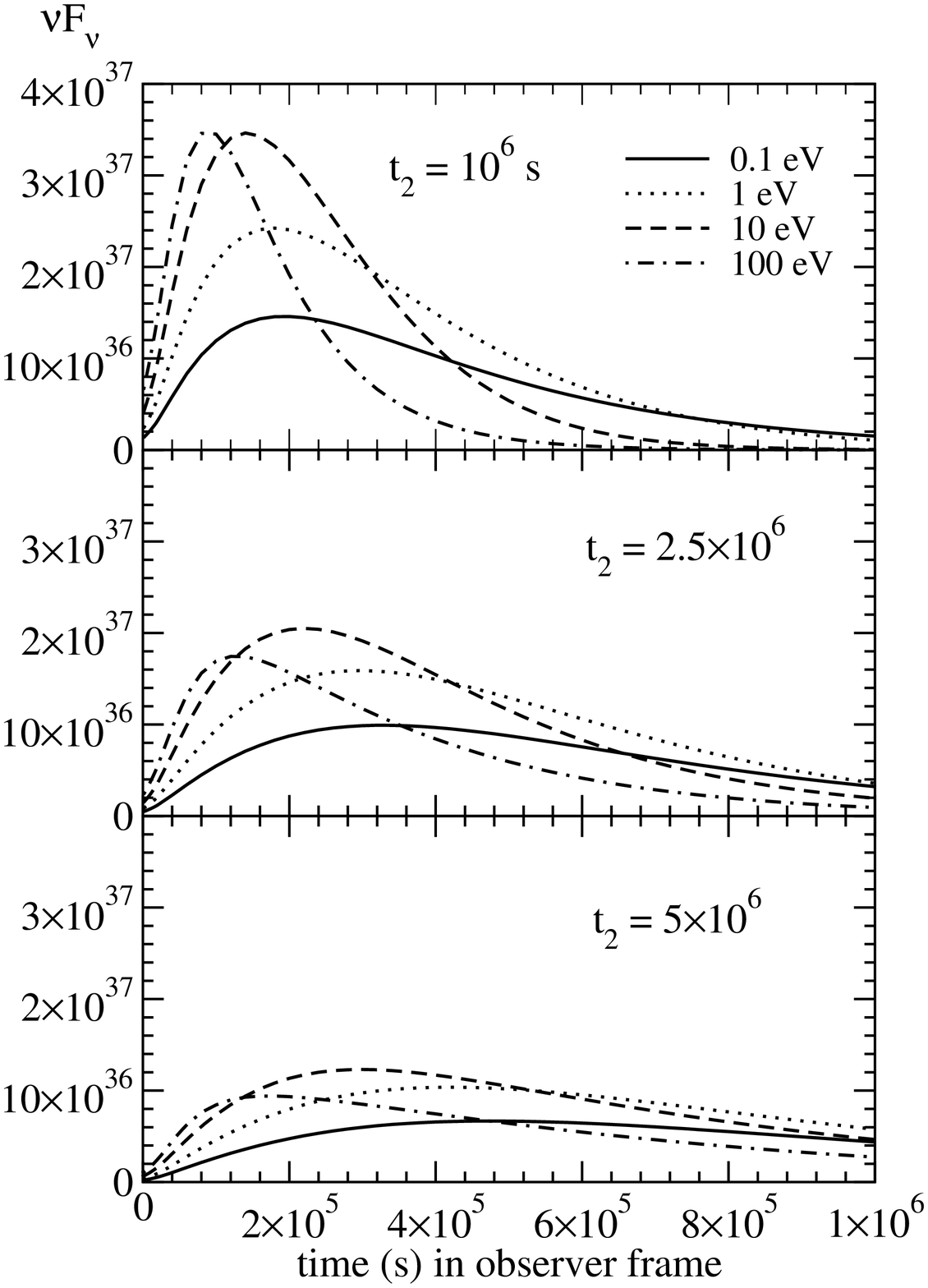}{190pt}{0}{45}{45}{-250}{-100}   
\plotfiddle{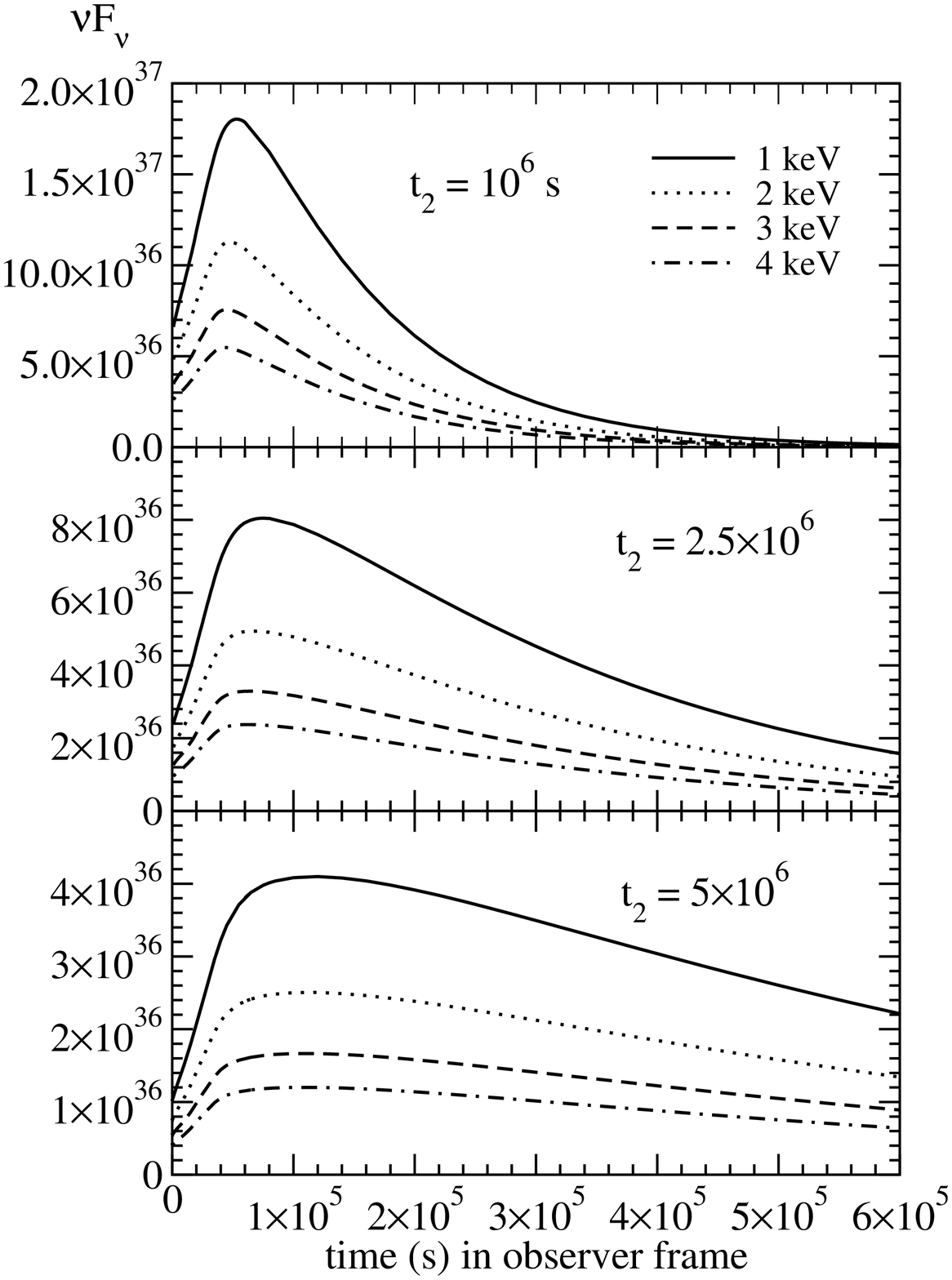}{190pt}{0}{45}{45}{10}{100}   
\vspace{0.1cm}   
\caption{The light curves for different injection timescales   
characterized by $t_2$ for a non-expanding blob:  
model E2, E1, and A1 from top to bottom.}   
\label{fig7}   
\end{figure}    
   
\newpage 
\begin{figure}   
\vspace{3.5cm}   
\epsscale{0.5}   
\plotfiddle{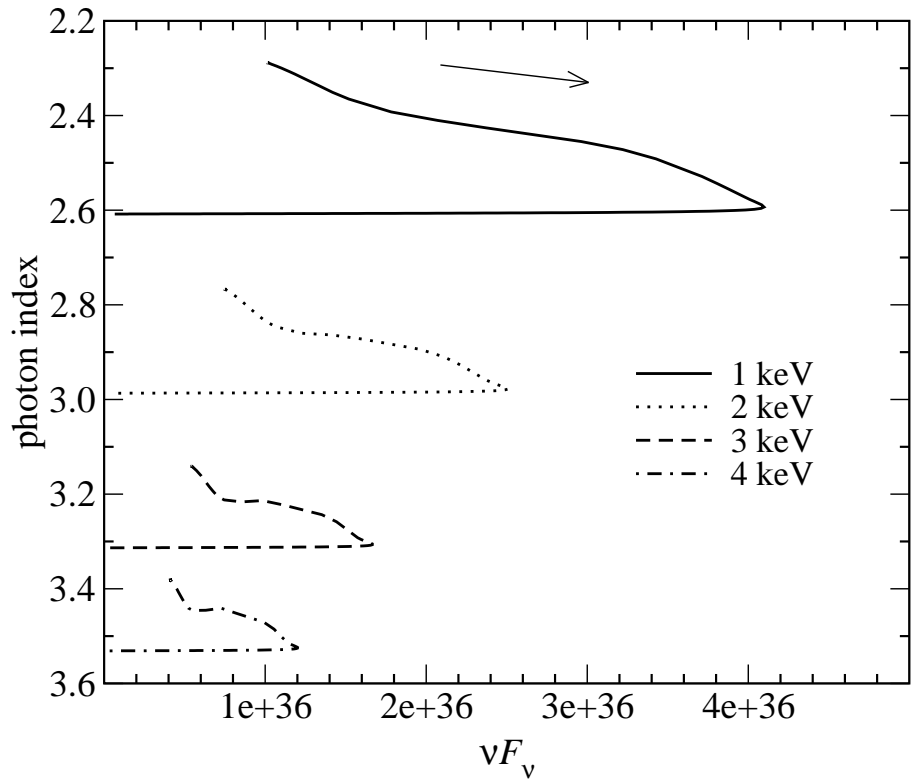}{190pt}{-90}{60}{60}{-250}{400}   
\vspace{0.1cm}   
\caption{The trajectory in the photon index $\alpha$ and  
luminosity plane for a non-expanding model A1.  The evolution is clockwise.}  
\label{fig8}   
\end{figure}    
   
\newpage 
\begin{figure}   
\vspace{3.5cm}   
\epsscale{0.5}   
\plotfiddle{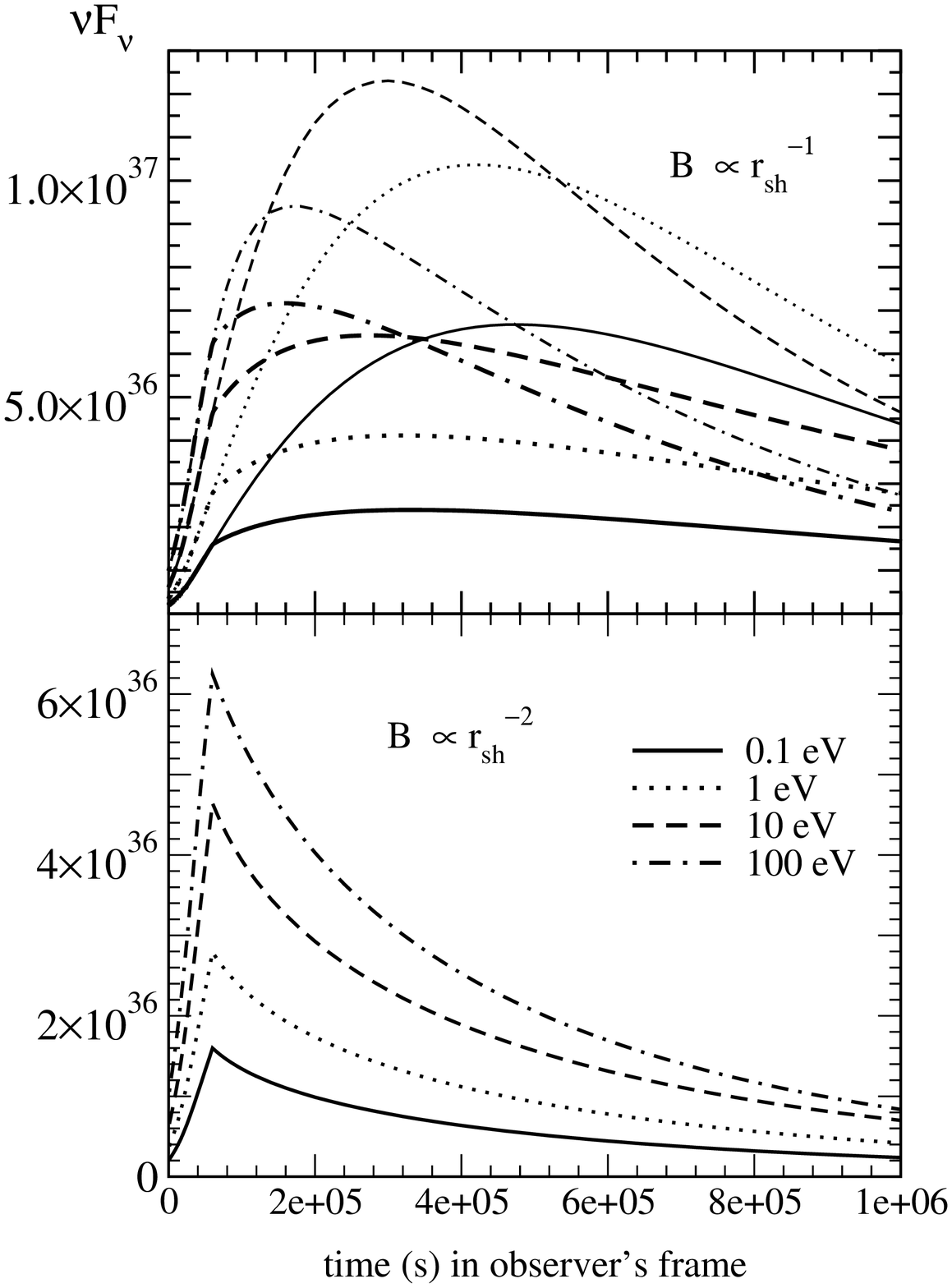}{190pt}{0}{45}{45}{-250}{-100}   
\plotfiddle{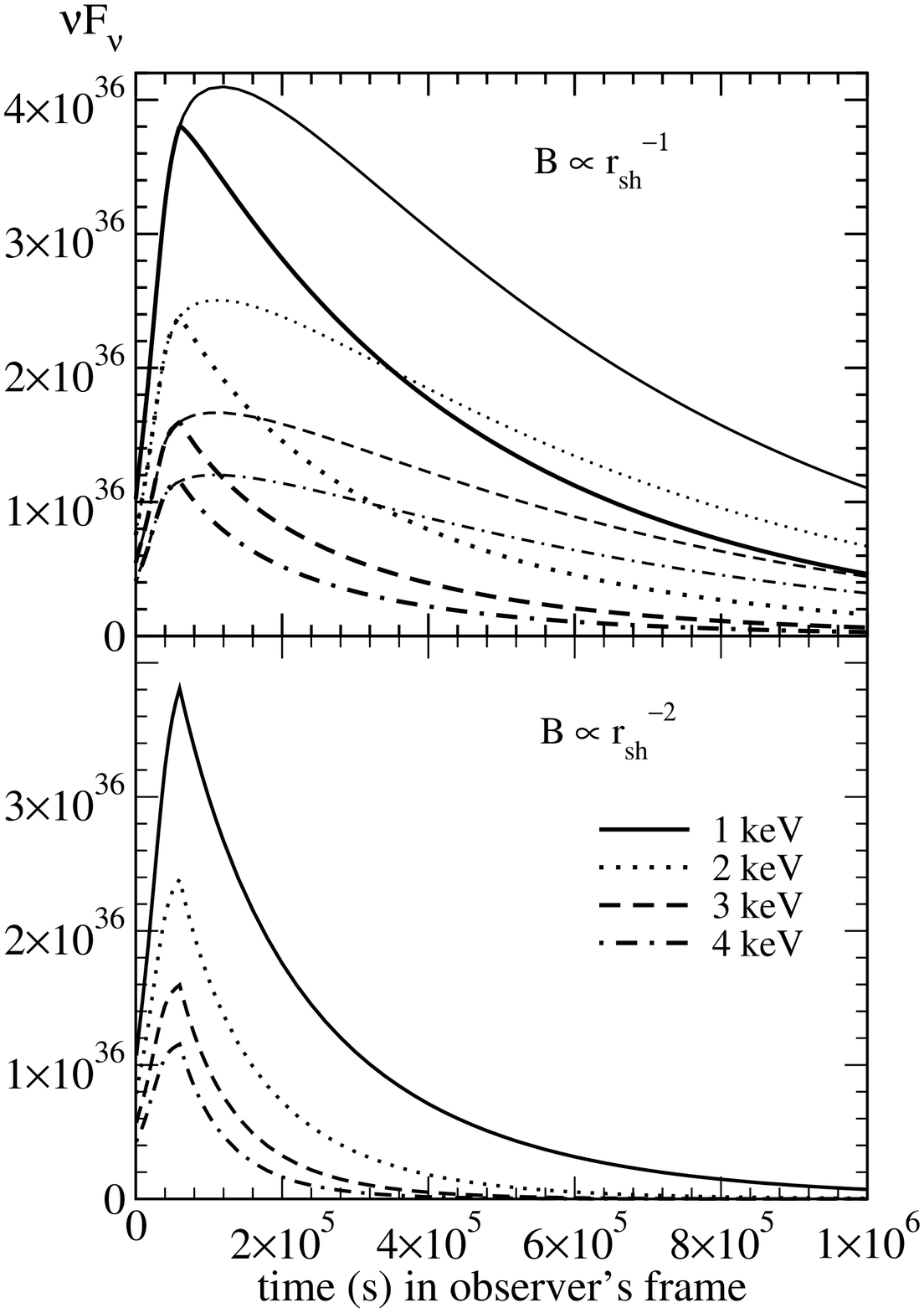}{190pt}{0}{45}{45}{10}{100}   
\vspace{0.1cm}   
\caption{The light curves for expansion cases: exp-C1 (top) exp-A1 (bottom).  
In the top panel, a non-expanding model A1 is also shown by thin curves  
for comparison.}   
\label{fig9}   
\end{figure}    
   
\newpage 
\begin{figure}   
\vspace{3.5cm}   
\epsscale{0.5}     
\plotfiddle{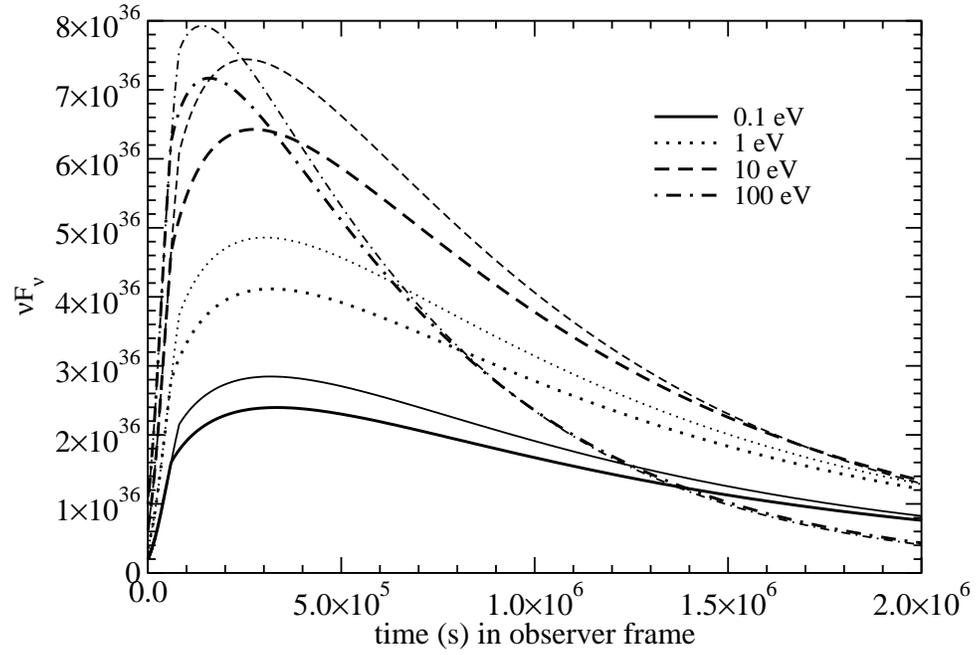}{190pt}{-90}{50}{50}{-200}{350}   
\plotfiddle{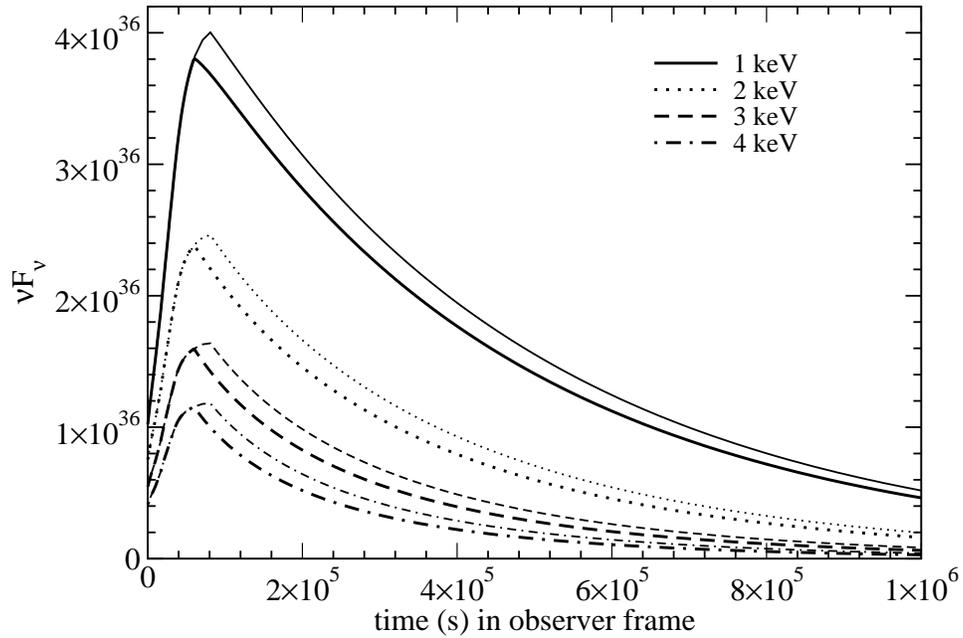}{190pt}{-90}{50}{50}{-200}{250}   
\vspace{1.0cm}   
\caption{The light curves for expansion cases with exp-C1 ($t_1 = 6 \times 10^5$ s)  
and exp-C2 ($t_1 = 8 \times 10^5$ s).  
Thick curves are for exp-C1 and thin curves are for exp-C2}   
\label{fig10}   
\end{figure}    
   
\newpage 
\begin{figure}   
\vspace{3.5cm}   
\epsscale{0.5}     
\plotfiddle{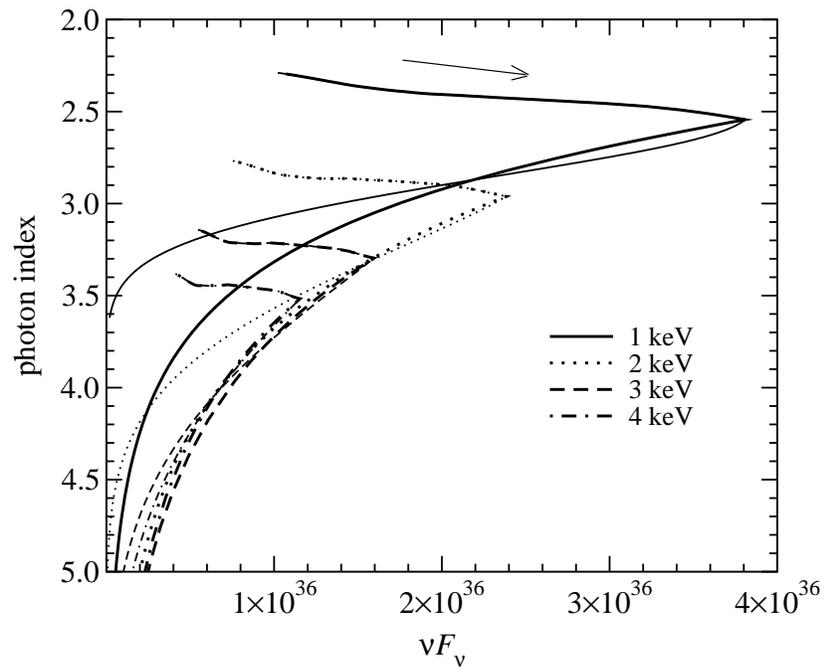}{190pt}{-90}{50}{50}{-200}{250}   
\vspace{1.0cm}   
\caption{The trajectory in the photon index and luminosity plane  
for expansion cases: model exp-A1 (thick curves) and exp-C1 (thin curves).}   
\label{fig11}   
\end{figure}    
   
\end{document}